\documentclass[a4paper]{JHEP3}
\pdfoutput=1
\usepackage{array}
\usepackage{epsfig}
\usepackage{amssymb}
\usepackage{slashed}
\usepackage{amsmath}

\usepackage{graphicx}
\graphicspath{{./Figs/}}
\usepackage{bm}

\newcommand{\<}{\langle}
\renewcommand{\>}{\rangle}
\newcommand{\beq}{\begin{equation}}
\newcommand{\eeq}{\end{equation}}
\newcommand{\beqn}{\begin{eqnarray}}
\newcommand{\eeqn}{\end{eqnarray}}
\newcommand{\al}{\alpha}
\newcommand{\D}{\Delta}
\newcommand{\eps}{\epsilon}
\def\sla#1{\setbox0=\hbox{$#1$}\dimen0=\wd0
      \setbox1=\hbox{/} \dimen1=\wd1 \ifdim\dimen0>\dimen1
      \rlap{\hbox to \dimen0{\hfil/\hfil}} #1                        \else
      \rlap{\hbox to \dimen1{\hfil$#1$\hfil}}
      /   \fi}

\newcommand{\nn}{\nonumber}
\newcommand{\ov}{\overline}
\newcommand{\mc}{\mathcal}

\newcommand{\id}{1 \hspace{1.15mm} \!\!\!\!1}

\preprint{%
TUM-HEP-732/09\\
CERN-PH-TH/2009-139\\
OHSTPY-HEP-T-09-001%
}

\title{Viable and testable SUSY GUTs with Yukawa unification: the case of split trilinears}

\author{Diego Guadagnoli$^{a,b}$, Stuart Raby$^c$ and David M. Straub$^a$\\
$^a$ Physik-Department, Technische Universit\"at M\"unchen, James-Franck-Str., 85748 Garching, Germany\\
$^b$ CERN, Theory Division, CH-1211 Geneva 23, Switzerland\\
$^c$ The Ohio State University, 191 W. Woodruff Ave., Columbus, OH 43210, USA\\
Email: \email{diego.guadagnoli@ph.tum.de, raby@mps.ohio-state.edu, david.straub@ph.tum.de}}

\abstract{
We explore general SUSY GUT models with exact third-generation Yukawa unification, but where the 
requirement of universal soft terms at the GUT scale is relaxed. We consider the scenario in which the breaking 
of universality inherits from the Yukawa couplings, i.e. is of minimal flavor violating (MFV) type. In particular, 
the MFV principle allows for a splitting between the up-type and the down-type soft trilinear couplings.

We explore the viability of this trilinear splitting scenario by means of a fitting procedure to electroweak 
observables, quark masses as well as flavor-changing neutral current processes. Phenomenological viability 
singles out one main scenario. This scenario is characterized by a sizable splitting between the trilinear soft terms 
and a large $\mu$ term. 
Remarkably, this scenario does not invoke a partial decoupling of the sparticle spectrum, as in the case of universal 
soft terms, but instead it {\em requires} part of the spectrum, notably the lightest stop, the gluino and the lightest 
charginos and neutralinos to be very close to the current experimental limits. The above mechanism is mostly triggered 
by a non-trivial interplay between the requirements of negative, sizable SUSY threshold corrections to $m_b$ and 
an instead negligible modification of the $B \to X_s \gamma$ decay rate, in presence of various other constraints, 
most notably a successful EWSB and a not too large BR($B_s \to \mu^+ \mu^-$).

We present a model-building interpretation of our discussed scenario and emphasize the crucial role of SUSY 
spectrum determinations at the LHC for either falsifying Yukawa unification or else providing important hints on the mechanism of 
SUSY breaking at work.
}


\begin{document}

\section{Introduction}

It remains a striking fact that the Standard Model (SM) gauge couplings, measured at low energies, 
evolve to a single value at a high scale $M_G$, provided that, above the electroweak scale, the SM 
is assumed to become supersymmetric. This fact may be an accident or may not be. In support of the 
second possibility are a few further remarkable features of this observed gauge coupling unification: 
it is very weakly dependent on the details of the supersymmetric spectrum, hence (presumably) a robust 
consequence of the assumed gauge symmetry and of supersymmetry (SUSY); furthermore, 
$M_G \approx 3 \times 10^{16}$ GeV happens to be at just the right place, namely above the region where 
proton decay at an unacceptable rate is generic, and below $M_{\rm Planck}$, where unavoidably large 
gravitational effects would make the calculation unrealistic. These considerations, together with the 
possibility to address various structural questions unanswered within the SM, warrant the decade-long 
interest in grand unified theories (GUTs). However, concerning further tests beyond that of gauge 
coupling unification, the possibility of general conclusions has been hampered by the larger number 
of model assumptions needed in each case. A prototype example is the already mentioned proton decay 
\cite{LucasRabyPRD,GotoNihei,BabuPatiWilczek,DermisekMafiRaby,Dutta04,Dutta07,Nath06,Nath07}.

An alternative way to test SUSY GUTs is by exploring the consequences of the generic expectation 
of low-energy SUSY. In general, the question of the predicted pattern of SUSY masses and mixings maps 
onto the question of the mechanism of SUSY breaking and of the form Yukawa couplings assume at the 
high scale. In turn, the latter issues usually require strong theoretical assumptions, that however may 
be justified, besides their possible appeal, if they lead to sufficiently sharp predictions.

A very elegant assumption, potentially testable in the SM fermion masses and mixings, is that of Yukawa 
unification (YU) at the GUT scale \cite{preHRS1,preHRS2,preHRS3,preHRS4,preHRS5,preHRS6,preHRS7,preHRS8}. 
It is motivated by the fact that, due to the higher degree of symmetry, matter fields must sit in 
appropriate representations of the gauge group, thereby sharing a common Yukawa coupling. Since 
this simple picture can be spoiled by e.g. the presence of higher-dimensional interactions, the 
crucial question is whether YU may leave any low-energy remnant at all. While for the light 
fermion generations this is definitely not the case,\footnote{%
Yukawa ratios different from 1, which are predicted in some GUT models, can however be phenomenologically viable \cite{Antusch:2008tf,Antusch:2009gu}.} 
for the third generation it remains an 
open and appealing possibility, selecting the group SO(10) as the potentially most predictive case. 
In this case, verification of the YU hypothesis would amount to the important conclusion that all 
the dimension-4 interactions involving 3rd-generation fermions and/or scalars originate from just 
the ${\mathbf{16_3~16_3~10_H}}$ structure. More shaky are symmetry assumptions, e.g. universality, 
on the soft-breaking terms, which correspond to operators of dimension less than 4, and are hence 
unlikely to preserve information about the symmetries inherent to the UV theory completion. The 
tacit motivation here is just one of computational simplicity.

In ref. \cite{AlGuRaS} the viability of the hypothesis of $t - b - \tau$ Yukawa unification 
in SUSY GUTs was studied, assuming that soft-breaking terms for sfermions and gauginos are universal 
at the GUT scale.
It was found that this hypothesis is challenged by the constraints imposed on the parameter space by 
FCNC processes, unless decoupling of the squark spectrum is invoked, 
thereby pushing the lightest squark well above 1 TeV. Under the same universality hypotheses, a viable 
alternative to decoupling has been found to be a moderate breaking of $t - b$ unification while keeping $b - \tau$
unification, or equivalently a parametric departure of $\tan \beta$ from the value implied by exact $t - b - \tau$ YU. 

The conclusions of ref. \cite{AlGuRaS} hold, we repeat, under the assumption, very common in 
the literature, of GUT-scale universalities for soft terms in the sfermion as well as in the gaugino sector.%
\footnote{On the other hand, soft terms for the Higgs scalars are allowed, and actually required, to be split 
from each other.}
An interesting question is then whether departures from GUT-scale universalities exist, that on the one 
hand allow Yukawa-unified SUSY GUTs to successfully withstand the combined constraints mentioned above 
without decoupling of the SUSY spectrum, and that on the other hand can be substantiated with a plausible 
SUSY-breaking mechanism.

In this respect, two simple and well-motivated scenarios of GUT-scale non-universalities 
emerge \cite{Snowmass96}: 
\begin{itemize}
\item Non-universal gaugino masses (NUGM), 
\item Non-universal scalar soft terms of minimally flavor-violating (MFV) form, i.e. inheriting from the SM 
Yukawa couplings.
\end{itemize}
These possibilities are not exclusive to each other -- e.g. there could even be a single $F$-term SUSY 
breaking causing non-universalities of both the above mentioned types at the same time \cite{Snowmass96}. 
However, to simplify matters, we will keep these two possibilities separate in the rest of the present 
discussion. In particular, since a detailed analysis of the NUGM scenario can be found in Baer {\em et al.}, 
ref. \cite{Baer-NUGM}, and a dedicated study for the case of YU in \cite{BalazsDermisek}, 
we will focus on the scenario of MFV soft terms.

Concretely, the purpose of the present paper is to address the following questions 
\begin{itemize}

\item[1.] whether relaxing GUT-scale universalities in favor of MFV soft terms allows, in the context of SUSY 
GUTs with YU, to recover phenomenological viability without invoking decoupling of the sfermion sector;

\item[2.] whether the experimental constraints used to address point 1 provide enough information 
to single out specific regions in the general parameterization, to be discussed below, that soft 
terms respect when they are of the MFV form.
\end{itemize}

The question of phenomenological viability, point 1, will be addressed by contrasting our 
class of models with established data on EW observables and flavor-changing neutral current (FCNC) 
decays. Our conclusions will be assessed through a fitting procedure, similar to that of refs. 
\cite{AABuGuS,AlGuRaS}, which has the advantage of being manifestly reparameterization-invariant. 
The details of this procedure will be presented in section \ref{sec:procedure}.

Turning to point 2, the corresponding question is relevant in connection with the search of a 
plausible mechanism of SUSY breaking able to substantiate the pattern of soft terms emerging from 
point 1. As anticipated at the end of the next section and detailed in our numerical analysis, section \ref{sec:results}, 
low-energy constraints are indeed powerful enough for a clear pattern of soft terms to emerge. 
A concrete example of a SUSY-breaking scenario where this pattern naturally emerges will be 
discussed in section \ref{sec:SUSYbreaking}.

Our results provide a concrete example where, under what we consider very plausible (and to our 
knowledge previously unexplored) assumptions for soft terms, enough remnant information on the 
high-energy symmetries survives at low energies for these symmetries to be reconstructible.
Our results also illustrate the crucial role that measurements of the lightest part of the SUSY 
spectrum play in this reconstruction program.

\section{Yukawa unification and the MFV principle}\label{sec:YU}

In order to introduce our problem of interest, let us first review briefly the case 
of Yukawa-unified SUSY GUTs where soft terms at the GUT scale are parameterized in terms of a 
universal soft mass $m_{16}$, a universal trilinear coupling $A_0$ and a universal gaugino mass
$m_{1/2}$.

The assumption of unification of the third generation Yukawa couplings at the scale 
$M_G\sim3\times 10^{16}$~GeV gives rise to two relations between the top quark, bottom quark and tau 
lepton masses. These relations depend on the renormalization group (RG) evolution, which is governed by 
gauge and Yukawa couplings only, and on weak scale threshold corrections, which depend on the soft 
SUSY breaking parameters, or, equivalently, on the sparticle spectrum and mixings. These threshold 
corrections are most important for the $b$ quark due to non-holomorphic contributions enhanced by 
$\tan\beta$ \cite{HRS}.
Neglecting these threshold corrections, requiring the tau lepton mass to equal 
its observed value of $1.777$~GeV and choosing $\tan\beta\sim50$ in order to reproduce the measured top 
quark mass \cite{CDFD0mt}, the running $b$ quark mass $m_b(m_b)$ would be predicted to be $4.5$~GeV, 
as opposed to the precisely measured experimental figure of $(4.20 \pm 0.07)$~GeV \cite{PDBook}. 
This illustrates the necessity of considering regions of SUSY parameter space where the overall threshold 
corrections to $m_b$ are {\em negative} \cite{BDR1,BDR2}. We will now discuss for which choices of GUT-scale 
parameters this is the case.

There are two dominant contributions to the threshold corrections to $m_b$, one arising from gluino-sbottom 
loops and one from chargino-stop loops. Considering only these two dominant contributions,\footnote{%
In the numerical analysis of section \ref{sec:results}, we take into account \emph{all} contributions 
to the threshold corrections.} the running $b$ quark mass at the decoupling scale can be written 
in terms of the running $b$ quark Yukawa coupling and the threshold corrections as \cite{HRS}
\beq
m_b = \frac{v y_b}{\sqrt{2}}\cos\beta \left(1 + \Delta_{\tilde g} + \Delta_{\tilde\chi}\right)~,
\eeq
where
\beqn
\Delta_{\tilde g} &=& \frac{2 g_3^2}{12\pi^2} \, \mu\tan\beta \, m_{\tilde g} \, 
I(m_{\tilde b_1}^2,m_{\tilde b_2}^2,m_{\tilde g}^2)~,
\label{eq:deltamb-g}
  \\
\Delta_{\tilde \chi} &=& \frac{y_t^2}{16\pi^2} \, \mu\tan\beta \, A_t \, 
I(m_{\tilde t_1}^2,m_{\tilde t_2}^2,\mu^2)~,
\label{eq:deltamb-ch}
\eeqn
and the loop function, which (for positive arguments) is strictly positive and has dimensions of inverse 
mass-squared, is
\beq
I(a,b,c) = - \frac{xy \ln x/y + yz \ln y/z + zx \ln z/x}{(x-y)(y-z)(z-x)}~.
\eeq

Taking the $\mu$ parameter to be positive, as is indicated by the muon $(g-2)$ anomaly, $\Delta_{\tilde g}$ 
leads to a positive correction of $m_b$, while the sign of $\Delta_{\tilde \chi}$ is given by the sign of 
the stop trilinear parameter $A_t$. The assumption of $\mu > 0$ will be kept throughout this paper as well.
To fulfill the condition $\Delta_{\tilde g}$+$\Delta_{\tilde \chi} < 0$, 
it is therefore necessary to have $A_t$ large and negative.\footnote{%
Our sign convention for $A_t$ is such that the off-diagonal entry of the tree-level stop mass matrix 
reads $m_t(A_t-\mu\cot\beta)$.}

Assuming a universal trilinear coupling $A_0$ and a universal gaugino mass $m_{1/2}$ at the GUT scale, 
the low-energy value of $A_t$ is given by
\beq
 A_t \approx -2.0 \, m_{1/2} + 0.23 \, A_0,
\label{eq:Atapprox}
\eeq
hence large negative $A_t$ requires large negative $A_0$ or large $m_{1/2}$. However, the latter possibility 
is precluded by the fact that the running gluino mass is given by
\beq
m_{\tilde g} = M_3 \approx 2.6 \, m_{1/2},
\eeq
hence small $m_{1/2}$ is required to suppress $\Delta_{\tilde g}$.\footnote{%
This tension can be relieved by allowing non-universal gaugino masses at the GUT scale. In that case 
eq.~(\ref{eq:Atapprox}) generalizes to $A_t \approx -0.2 M_2-1.8 M_3 + 0.23 A_0$. As mentioned in the introduction, 
this NUGM scenario can lead to viable models of YU, but will not be considered in the following.}

Even with $|A_t| \gg m_{1/2}$, the gluino contribution to $m_b$ is still competitive with the chargino contribution 
in a large portion of MSSM parameter space. This is why an additional suppression is necessary, which can be achieved 
by a large stop-sbottom mass splitting, so regions of parameter space where the stop is the lightest sfermion are preferred.

All the features described above can be realized in the framework of the NUHM, the MSSM with non-universal Higgs mass 
parameters, and indeed in refs. \cite{BDR1,BDR2} (see also \cite{BF}) the region with
\beq
-A_0\approx 2\,m_{16}, ~~ \mu,m_{1/2} \ll m_{16},
\label{eq:ISMH}
\eeq
was found to allow successful YU. Within the parameter space of eq. (\ref{eq:ISMH}), successful EWSB
requires Higgs mass-squared parameters with the pattern $m_{16}^2 < m_{H_u}^2 < m_{H_d}^2$.

However, relations (\ref{eq:ISMH}), together with the large value of $\tan \beta \approx 50$ required for 
YU, have an important impact on the SUSY spectrum and on the predictions for FCNCs, in particular 
on those $B$-physics decay modes that are especially sensitive to large $\tan \beta$ and to the 
large trilinear coupling $A_t$ implied by relations (\ref{eq:ISMH}). Specifically, the decay modes 
that turn out to have the strongest impact are $B_s \to \mu^+ \mu^-$, $B \to X_s \gamma$ and 
$B \to X_s \ell^+ \ell^-$. For example, in the parameter space (\ref{eq:ISMH}), one typically has to
face a substantial enhancement of BR($B_s \to \mu^+ \mu^-$) and huge destructive interference from 
chargino contributions in BR($B \to X_s \gamma$).

In ref. \cite{AlGuRaS} the non-trivial interplay among these observables and the bottom mass 
has been studied extensively through a fitting procedure. The main conclusions were: 
\begin{itemize}
\item[{\em (a)}] that a generic SUSY GUT with exact YU and GUT-scale universalities for sfermions 
and gauginos is phenomenologically viable only by advocating partial decoupling of the sfermion 
sector, the lightest mass exceeding 1 TeV; 
\item[{\em (b)}] that phenomenological viability can be recovered without decoupling by 
relaxing $t-b-\tau$ unification to $b -\tau$ unification, equivalent to a parametric departure 
of $\tan \beta$ from the value implied by exact YU. This solution is non-trivial since, while 
the FCNC constraints prefer lower values of $\tan \beta$, a successful prediction of $m_b$ in 
YU requires instead a value of $\tan \beta$ very close to 50 \cite{CPW,ananthanarayan}. Indeed, a compromise solution 
between the two classes of constraints has been found to exist only for the narrow range 
$46 \lesssim \tan \beta \lesssim 48$, implying that the breaking of $t - b$ YU must be 
{\em moderate}, in the range 10 -- 20\%.
\end{itemize}

As stated in the introduction, our aim here is to address instead whether the problem in 
point {\em (a)} can be reconciled with the assumption of {\em exact} $t-b-\tau$ Yukawa unification by
relaxing instead the strong (and theoretically poorly justified) hypothesis of GUT-scale universalities 
in soft terms.

We will focus in this paper on non-universal soft terms for scalars satisfying the principle of minimal 
flavor violation (MFV) \cite{MFV,GrinsteinMFV}, corresponding to the assumption that the threshold at which the flavor 
symmetry is broken lies above the scale at which the soft terms are specified, and that the only spurions 
of the broken flavor symmetry are the Yukawa couplings of the corresponding SM interactions, so that the 
soft terms flavor structure must inherit from the SM Yukawas themselves. Then, restricting to the partners 
of the quark sector, soft scalar mass and trilinear terms have the form \cite{Snowmass96,MFV}
\beqn
m_Q^2 &=& \ov m_Q^2 (\id + c^u_Q Y_U Y_U^\dagger +c^d_Q Y_D Y_D^\dagger + O(Y_{U,D}^4))~, \nn \\
m_U^2 &=& \ov m_U^2 (\id + c^u_U Y_U^\dagger Y_U + O(Y_{U}^4))~, \nn \\
m_D^2 &=& \ov m_D^2 (\id + c^d_D Y_D^\dagger Y_D + O(Y_{D}^4))~, \nn \\
A_U &=& \ov A_U Y_U (\id + O(Y_{D}^2))~, \nn \\
A_D &=& \ov A_D Y_D (\id + O(Y_{U}^2))~,
\label{eq:softMFV}
\eeqn
where the $\ov m$ and $\ov A$ parameters have mass dimension 1 and the $c$ parameters are real, $O(1)$ numbers.

The hypothesis of exact YU --~and use of the hierarchical structure of Yukawa couplings~-- 
allows to drastically simplify expansions (\ref{eq:softMFV}). The SUSY-breaking terms in eq. 
(\ref{eq:softMFV}) assume in fact the approximate pattern
\beqn
 m_{Q,U,D}^2  \simeq
\begin{pmatrix}
\ov m_{Q,U,D}^2 & 0 & 0 \\
0 & \ov m_{Q,U,D}^2 & 0 \\
0 & 0 & \ov m_{Q,U,D}^2 + \Delta m_{Q,U,D}^2
\end{pmatrix} \,,
\label{eq:softMFVYU-m2}
\eeqn
\beqn
 A_{U (D)}  \simeq
\begin{pmatrix}
~0~ & ~0~ & ~0~ \\
~0~ & ~0~ & ~0~ \\
~0~ & ~0~ & ~y_{t (b)} \ov A_{U (D)}~
\end{pmatrix} \,,
\label{eq:softMFVYU-A}
\eeqn
i.e. they can be taken as diagonal and split between the third and the first two generations.\footnote{%
To establish contact with $A_{t,b}$, used before in the text, we note that $(A_{U,D})_{33} = y_{t,b} A_{t,b}$.}
This approximation is phenomenologically valid up to terms of the order $(Y_{U,D}^2)_{ij} / y^2_{33}$ 
with $i,j \neq 3$ and $y_{33}$ the common value of the GUT-scale Yukawa coupling for the third 
generation fermions. In particular, this approximation is sufficient to reproduce the most important 
features of the low-energy sparticle spectrum, while the neglected off-diagonal terms are of subleading 
importance for the spectrum, and also of subleading importance for the success of YU.

Therefore, in the instance where the SM flavor symmetry group is broken minimally at a scale 
higher than $M_{\rm GUT}$, the hierarchy of Yukawa couplings and the assumption of YU allow 
to parameterize the GUT-scale soft SUSY-breaking terms in the squark sector in a generic way with 6 real parameters 
for bilinear soft terms and 2 complex parameters for trilinear soft terms.

Given this general parameterization of squark soft terms, three scenarios of non-universalities compatible 
with the MFV principle suggest themselves:
\begin{enumerate}
\item generational bilinear splitting, $\Delta m_{Q,U,D}^2 \neq 0$ ,
\item up-down bilinear splitting, $\ov m_{Q}^2 \neq \ov m_{U}^2 \neq \ov m_{D}^2$ ,
\item up-down trilinear splitting, $\ov A_U \neq \ov A_D$ .
\end{enumerate}

Concerning the first of these possibilities, it is interesting to note that the viability of YU, as described 
at the beginning of this section, is essentially determined by quantities related to the third squark generation, 
such as the stop and sbottom spectrum; furthermore, conditions (\ref{eq:ISMH}), which were found to be favourable 
to YU, naturally lead to a large hierarchy between light third generation squarks and heavy first and 
second generation squarks, i.e. to an inverse scalar mass hierarchy \cite{Bagger:1999sy}. Thus, changing the masses of first 
and second generation squarks by means of generational bilinear splitting is not expected to have a strong 
impact on the success of YU. Indeed, our initial numerical explorations of YU in this scenario pointed to preferred 
values of $\Delta m_{Q,U,D}^2 \approx 0$. Hence, we will not consider any generational splitting in the 
following. We note however that it might help accommodate additional constraints, like the dark matter relic density, 
while not upsetting the mechanism ensuring the success of YU \cite{Auto:2004km}.

Scenario 2. instead can have a more profound impact on YU, since the different GUT-scale values for the up-type and 
down-type squarks can lead to a larger hierarchy between the stop and sbottom masses at low energies than is possible 
in the universal case. This would in turn allow a suppression of the unwanted positive gluino contributions to $m_b$. Indeed, 
such scenario has been studied before in the context of $b-\tau$ unification \cite{Komine:2001rm,Pallis:2003aw,Profumo:2003ema}. 
Whether it is possible to accommodate full $t-b-\tau$ YU and to satisfy all FCNC constraints within this scenario would 
require a dedicated analysis, which we however leave to a future study.

For the remainder of this work, we will thus concentrate on the study of the {\em trilinear splitting} scenario, 
where by definition a splitting between the up-type and down-type trilinear couplings is assumed, whereas sfermion 
bilinears are still taken as universal. That is, our assumptions for the soft terms of squarks and sleptons 
at the GUT scale are
\beqn
m^2_{Q,U,D,L,E} = m_{16}^2 \id \,,
\label{eq:ST-m2}
\eeqn
\beqn
A_U = \ov A_U Y_U \,, \qquad
A_D = \ov A_D Y_D \,, \qquad
A_L = \ov A_D Y_L \,.
\label{eq:ST-A}
\eeqn

\section{Procedure}\label{sec:procedure}

Our problem of interest is to study the MFV scenario with soft terms defined by eqs. (\ref{eq:ST-m2})--(\ref{eq:ST-A}) 
and a positive $\mu$ parameter
in the framework of a generic SUSY GUT with exact $t - b - \tau$ YU. The present section is devoted to the discussion 
of the procedure adopted in our numerical analysis.

\TABLE[ht]{
\begin{tabular}{|lc|lc|}
\hline
Observable & ~Value($\sigma_{\rm exp}$)~ & Observable & ~Lower Bound~ \\
\hline
$M_W$ & $80.398(25)$ & $M_{h^0}$ & $114.4$ \\
$M_Z$ & $91.1876(21)$ & $M_{\tilde \chi^+}$ & $104$ \\
$10^{5} G_\mu$ & $1.16637(1)$ & $M_{\tilde t}$ & $95.7$ \\
$1/\al_\text{em}$ & $137.036(0)$ & & \\
$\al_s(M_Z)$ & $0.1176(20)$ & & \\
$M_t$ & $173.1(1.3)$ & & \\
$m_b(m_b)$ & $4.20(7)$ & & \\
$M_\tau$ & $1.777(0)$ & & \\
\hline
\end{tabular}
\caption{Flavor conserving observables \cite{PDBook,CDFD0mt} used in the fit. 
Dimensionful quantities are expressed in powers of GeV.}
\label{tab:obs-EW}
}

\TABLE[t]{
\begin{tabular}{|lcr|}
\hline
Observable & ~~Value($\sigma_{\rm exp}$)($\sigma_{\rm theo}$)~~ & Ref. \\
\hline
$\D M_s / \D M_d$ & 35.1(0.4)(3.6) & \cite{Barberio08,CDF-DMs} \\
$10^4$ BR$(B \to X_s \gamma)$ & 3.52(25)(46) & \cite{Barberio08} \\
$10^6$ BR$(B \to X_s \ell^+ \ell^-)$ & 1.60(51)(40) & \cite{Babar-bsll,Belle-bsll} \\
$10^4$ BR$(B^+ \to \tau^+ \nu)$ & 1.40(40)(26) & \cite{PDBook} \\
BR$(B_s \to \mu^+ \mu^-)$ & $< 5.8 \times 10^{-8}$ & \cite{Bsmumu-CDF} \\
\hline
\end{tabular}
\caption{Flavor-changing observables used in the fit. The BR$(B \to X_s \ell^+ \ell^-)$ 
is intended in the range $q^2_{\ell^+ \ell^-} \in [1,6]$ GeV$^2$.
}
\label{tab:obs-FC}
}

The parameter space of our considered class of models is constrained through a fitting procedure against 
low-energy observables, that are reported in tables \ref{tab:obs-EW} and \ref{tab:obs-FC} along with their 
current experimental determinations. Specifically, a quantitative test of the model is obtained through a $\chi^2$ 
function defined as
\beqn
\chi^2[\vec\vartheta] \equiv \sum_{i = 1}^{N_{\rm obs}}
\frac{(f_i[\vec\vartheta] -\mc O_i)^2}
{(\sigma_i^2)_{\rm exp} + (\sigma_i^2)_{\rm theo}}~,
\label{eq:chi2}
\eeqn
where $\mc O_i$ indicates the experimental value of the observables and $f_i[\vec \vartheta]$ the corresponding 
theoretical prediction, which will be function of the model parameters
collectively indicated with $\vec \vartheta$. For each quantity, the experimental and theoretical 
standard deviations are also reported in tables \ref{tab:obs-EW} and \ref{tab:obs-FC}. For those
among the observables having a negligible experimental error, we took as overall uncertainly 
0.5\% of the experimental value, which we consider a realistic estimate of the numerical error 
associated with the calculations. For the details of the estimation of theoretical errors on the 
flavor observables (table \ref{tab:obs-FC}), we refer the reader to the comments reported in ref. 
\cite{AlGuRaS}.

In evaluating the $\chi^2$ function, we also included the bounds reported in tables \ref{tab:obs-EW} and 
\ref{tab:obs-FC}, in the form of suitably smoothened step functions added to the $\chi^2$ function and returning 
zero in the case of a respected constraint. Concerning SUSY masses, the bounds used in our analysis are 
the most conservative ones reported by the PDG \cite{PDBook}. In particular, we do not use an explicit bound
on the gluino mass, since that on the lightest chargino turns out to be strong enough. On the SUSY mass bounds 
used we will comment again in section \ref{sec:spectrum}.

The $\chi^2$ function is minimized using {\tt MIGRAD}, which is part of the {\tt CERNlib} library \cite{CERNlib}. 
The minimization procedure guarantees the invariance of our conclusions under reparameterizations of the theory.

\TABLE[ht]{
\renewcommand{\arraystretch}{1.3}
\begin{tabular}{|lcc|}
\hline
Sector & ~\#~ & Parameters \\
\hline \hline
gauge & 3 & $\alpha_G$, $M_G$, $\eps_3$ \\
SUSY-breaking & 6 & $m_{16}$, $m_{1/2}$, $m_{H_u}$, $m_{H_d}$, $\ov A_U$, $\ov A_D$ \\
SUSY (EW scale) & 2 & $\tan \beta$, $\mu$ \\
neutrino & 1 & $M_{R}$ \\
3rd generation Yukawa & 1 & $y_t=y_b=y_\tau=y_{\nu_\tau}$ \\
\hline
light generation Yukawa & 6 & $y_{u,c}$, $y_{d,s}$, $y_{e,\mu}$ \\
CKM & 4 & $\lambda$, $A$, $\bar\varrho$, $\bar\eta$ \\
\hline
\end{tabular}
\renewcommand{\arraystretch}{1}
\caption{Model parameters. Unless explicitly stated, they are intended at the GUT scale.}
\label{tab:parameters}
}

The full set of free parameters describing our considered class of models is collected in 
table~\ref{tab:parameters}.
For our purposes, we can take the grand unified group, on which we do not have to make any assumptions, to be broken 
to the SM group in one single step beneath the unification scale $M_G$.
We allow for a percent level threshold correction $\epsilon_3$ to the strong gauge coupling at the GUT scale. 
Our assumption of universal sfermion masses and split (but real) trilinear couplings amounts to six parameters in the soft 
SUSY breaking sector. The MSSM renormalization group equations (RGEs) for the GUT-scale parameters are solved between 
$M_G$ and the EW scale, where we define $\mu$ and $\tan\beta$.

In order to account, in this RG evolution, for the (possible) effects of right-handed neutrinos, 
present e.g. in SO(10) and required for the see-saw mechanism, we allow for the contribution of a 
third-generation neutrino Yukawa coupling (with initial condition $y_{\nu_\tau} = y_t$) in all RGEs 
between $M_G$ and the right-handed neutrino threshold $M_R < M_G$ \cite{Hisano95,AKLR02,Petcov03}.
We stress that, since we are not considering any particular model construction and are only concerned with the phenomenological
viability of Yukawa unification, it is sufficient in our setup to consider only a single right-handed neutrino threshold 
and the details of the neutrino Yukawa texture are not relevant. Therefore, we simply assume 
$(Y_\nu)_{ij}=y_t \delta_{i3}\delta_{j3}$ at $M_G$. The concrete impact of the inclusion of $M_R$ on our results will be 
discussed in section~\ref{sec:RHnu}.

It is worth stressing that the inclusion of the RH 
neutrino scale in the running as discussed above allows to eliminate a potential source of large logarithmic GUT-scale 
threshold corrections to YU as well as to Higgs splitting.
With the above said, it is difficult to exclude, in our fully general approach, the presence of 
additional GUT-scale threshold corrections to the Yukawa couplings. However, since our aim is to explore 
the predictive power of the hypothesis of YU, we will assume residual threshold corrections to be negligible, 
i.e. to leave YU a well defined hypothesis at the GUT scale. This assumption is realized in many concrete 
models, see e.g. \cite{BDR2}.

Our treatment of the RGE running, inclusion of threshold corrections, determination of the Higgs 
VEVs and calculation of the various observables is largely similar to that of Refs. \cite{AABuGuS,AlGuRaS}.
Here we just spell out a few improvements,  namely: the use of two-loop RGEs for the soft sector, the 
Yukawa and the gauge couplings \cite{MartinVaughn}; the inclusion of the full $3\times3$ flavor dependence
of the Yukawa couplings (in place of neglecting the effect of Yukawa entries different from the 33 
one). Concerning the latter point, while this in principle introduces 10 additional free parameters into 
the fitting procedure (cf. table~\ref{tab:parameters}), it does not pose a problem in practice (e.g. of fitting convergence).
Indeed, the light Yukawa couplings and CKM entries are largely insensitive to changes in the SUSY parameters
and, given also their hierarchical structure, can be determined very easily for each fit point. 
We further note that, while this approach is in principle valid for any Yukawa texture specified 
at $M_G$, we assumed exact YU to hold for the 33 elements of the Yukawa matrices in the basis where the down-type 
quark and charged lepton Yukawas are diagonal. In models predicting large 23 or 32 elements in Yukawa matrices, 
there would be small corrections to exact YU in this basis.

Before concluding this section, we would like to add a further general comment on the procedure described
above. It is clear that addressing the viability of the YU hypothesis is a non-trivial task also 
on purely computational grounds, since, in addition to the RG evolution of the relevant couplings, one has 
to take into account also threshold corrections to fermion masses, that depend on the details of the SUSY 
spectrum, as well as uncertainties in the measured values of fermion masses themselves and additional 
constraints on the parameter space, like FCNC processes. To deal with these problems, two different approaches 
have been followed in the literature.

The first approach \cite{BalazsDermisek,Tobe:2003bc,Auto:2003ys,Baer:2008xc,Antusch:2008tf,Antusch:2009gu,Baer:2008jn,Baer:2008yd,Gogoladze} 
amounts to fixing the low-scale values of the fermion masses to their experimental central values. 
After taking into account weak-scale threshold corrections and the running of the Yukawa couplings to 
the GUT scale, the differences between $y_t$, $y_b$ and $y_\tau$ as functions of the SUSY breaking 
parameters quantify the quality of the -- approximate -- YU. Apart from computational simplicity, 
this approach is motivated by the fact that exact YU might be spoiled by higher-dimensional operators 
and it allows to sacrifice some amount of unification to accommodate additional constraints, like the 
dark matter relic density.

The second approach \cite{BDR1,BDR2,AABuGuS,AlGuRaS}, which we adhere to, amounts to imposing exact 
YU at the GUT scale, while fitting the low-energy values of fermion masses to their observed values by 
means of a $\chi^2$ minimization procedure. In our view, in spite of possibly being computationally more demanding, this approach has several 
advantages: First, the fitting procedure automatically singles out regions in parameter space favouring 
YU, while these regions have to be found by a widespread scan in the previous approach; second, the 
exact YU case represents a benchmark case exploiting the maximal predictive power of the YU hypothesis, 
hence leading to clear-cut predictions about qualitative features of the SUSY spectrum, which are univocally 
falsifiable at the LHC. This predictability has the downside that e.g. it may not be as easy to reconcile 
the standard neutralino relic density with YU in this scenario. However, we note that this constraint is a very 
indirect one, since it can easily be circumvented by departing from standard cosmology, without 
compromising the other ingredients of the approach.

\section{Numerical analysis}\label{sec:results}

In this section we would like to show that the trilinear splitting scenario, as defined at the end of section \ref{sec:YU},
allows generic SUSY GUTs with {\em exact} YU to recover full phenomenological viability without invoking 
decoupling, at variance with what happens in the case of universal trilinears \cite{AABuGuS,
AlGuRaS}. This fact is due to a non-trivial interplay between the $\mu$ parameter and the $\ov A_U - \ov A_D$ 
splitting. The basic picture is that these 
parameters control the SUSY corrections to the bottom mass $m_b$ and to the $B \to X_s \gamma$ decay
rate, the latter representing the two main observables that generate a problem in the case of universal 
soft terms, once other constraints, notably BR($B_s \to \mu^+ \mu^-$) and BR($B \to X_s \ell^+ \ell^-$), 
are taken into account.
In the numerical analysis, we restricted ourselves to a fixed universal sfermion mass $m_{16}$ of 4 TeV, 
also for the sake of comparison with refs. \cite{AABuGuS,AlGuRaS}. While a change in this parameter would 
change the allowed room for the stop-sbottom splitting mentioned in section~\ref{sec:YU}, it would leave 
the lightest part of the spectrum, which is relevant in particular for collider phenomenology, unaffected 
as we will see.

In the following subsection, we will present the main results of our numerical analysis, and their theoretical interpretation
will be provided in section \ref{sec:interpretation}. Sections \ref{sec:spectrum}-\ref{sec:gm2} are devoted to additional considerations, 
related to the implied sparticle mass spectrum, to the role of right-handed neutrinos, and to the $(g-2)_\mu$ and 
dark matter constraints.

\FIGURE[ht]{
\includegraphics[width=0.45\textwidth]{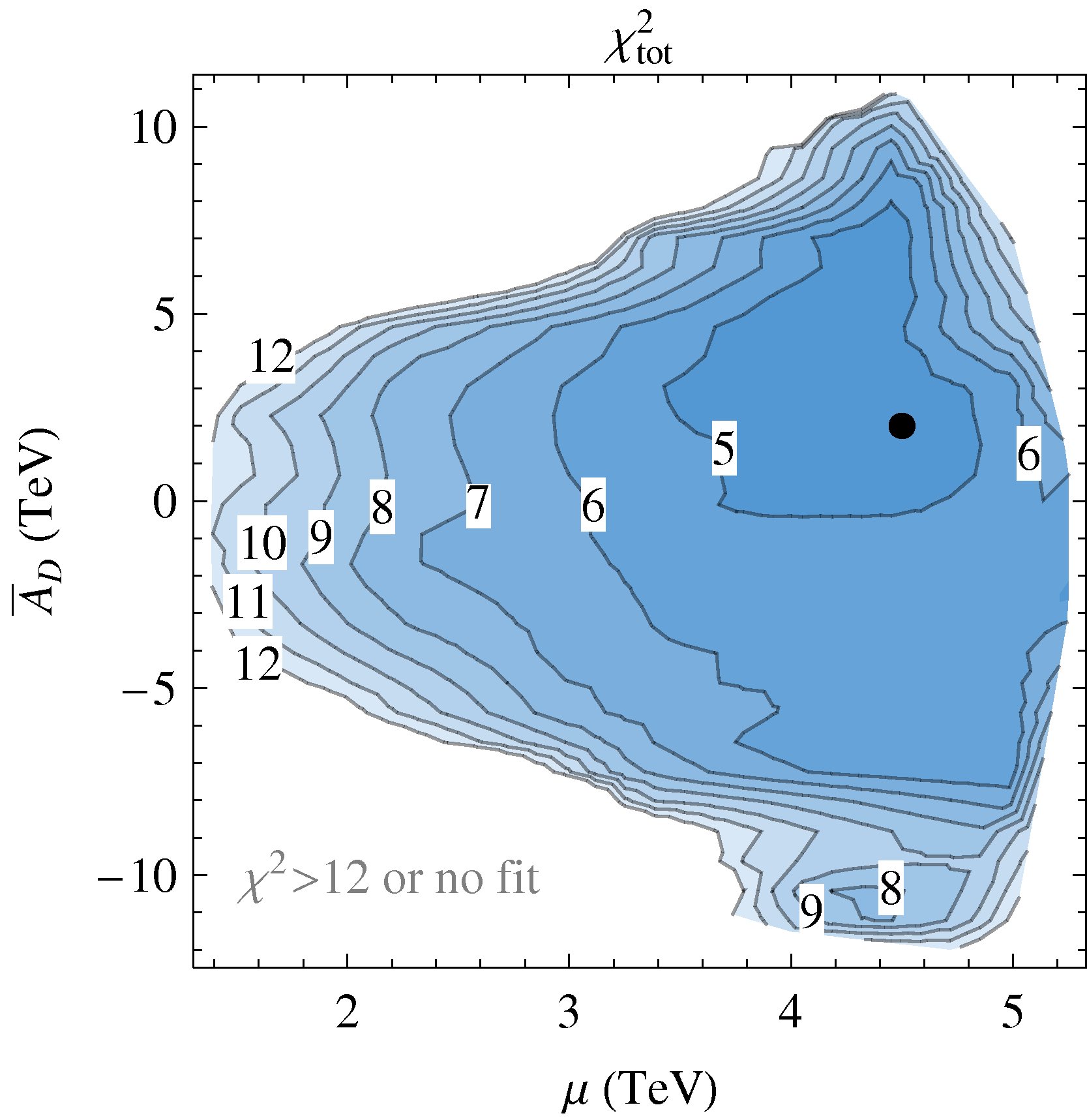}
\hspace{0.04\textwidth}
\includegraphics[width=0.45\textwidth]{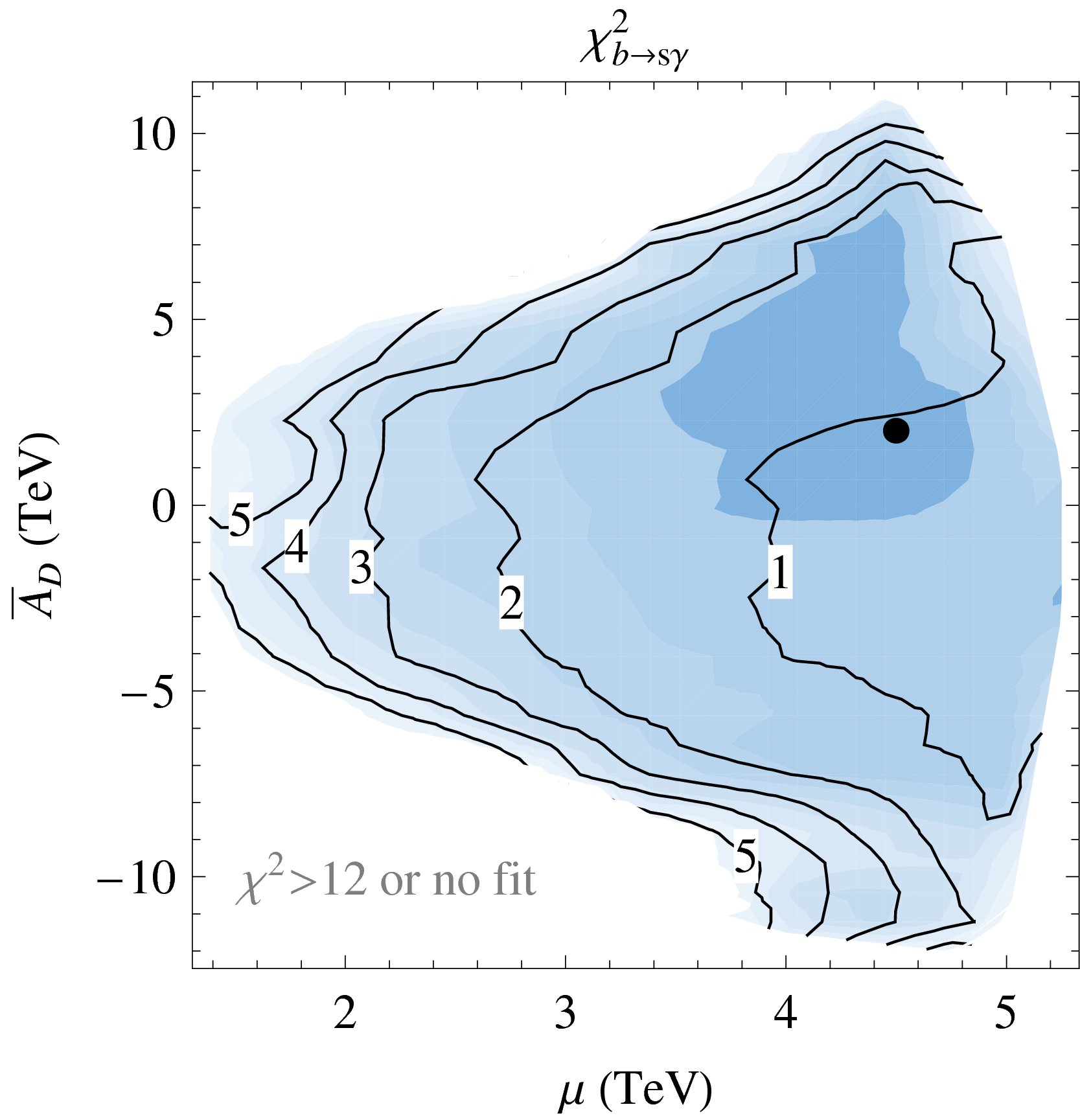}
\caption{Left: lines of constant $\chi^2$ in the $\mu$ vs $\ov A_D$ plane, with $m_{16} = 4$ TeV. 
Right: contributions to the $\chi^2$ function from $\text{BR}(B\to X_s\gamma)$. The black dot 
represents the example fit reported in tables~\ref{tab:examplefit}-\ref{tab:examplefit-par}.}
\label{fig:contours-chi2}
}

\FIGURE[ht]{
\includegraphics[width=0.45\textwidth]{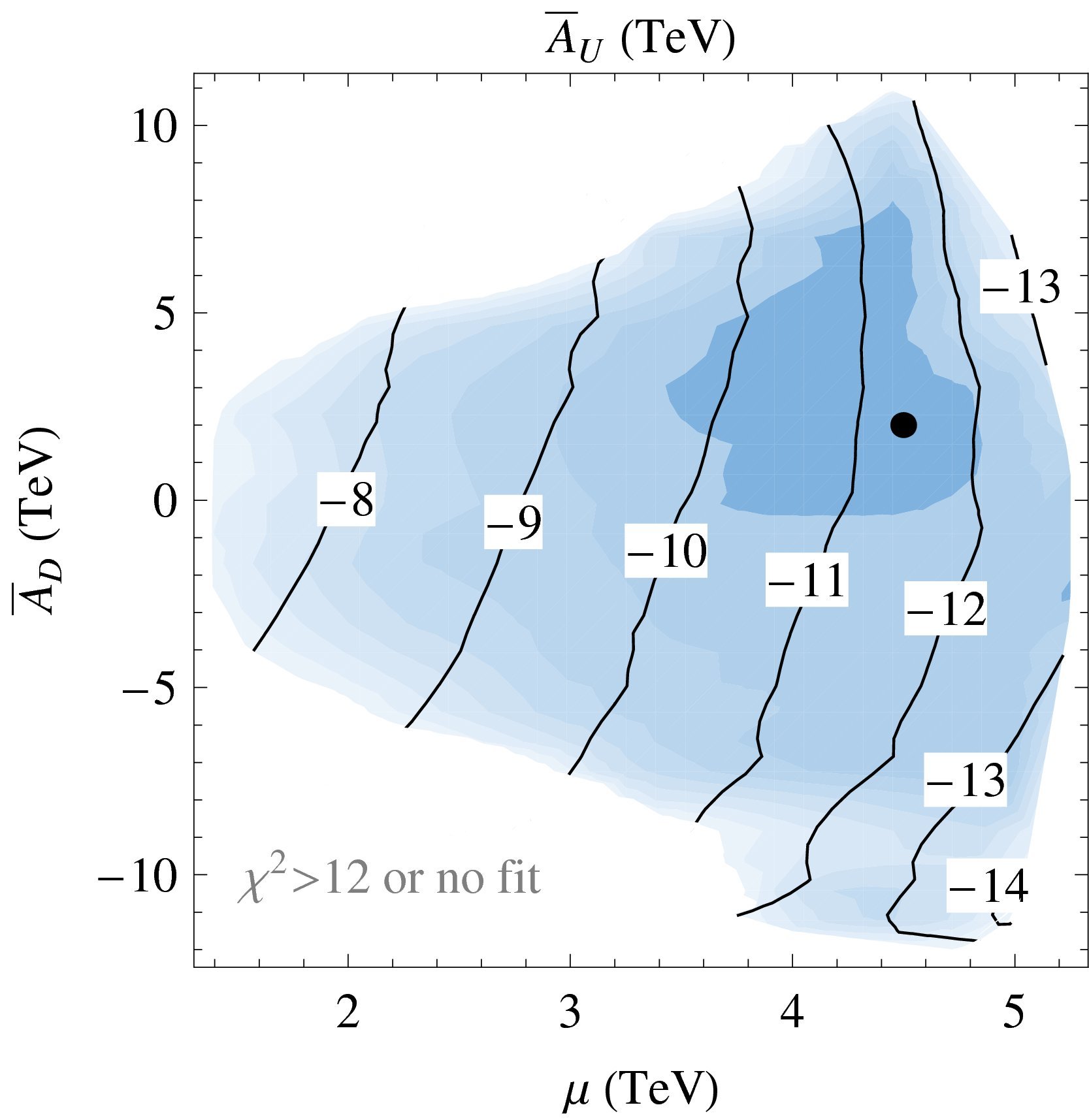}
\hspace{0.04\textwidth}
\includegraphics[width=0.45\textwidth]{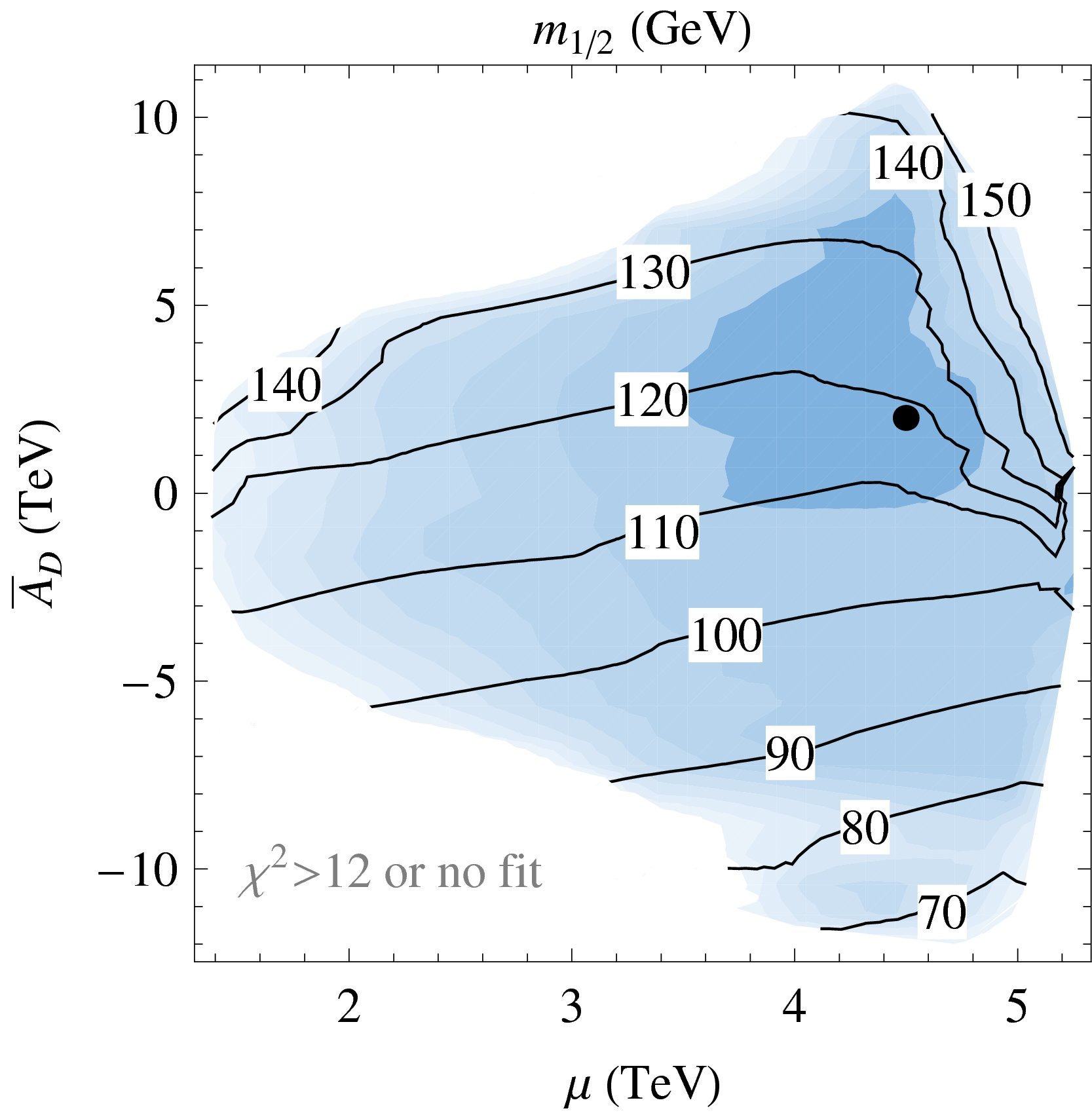} \vspace{0.2cm}\\
\includegraphics[width=0.45\textwidth]{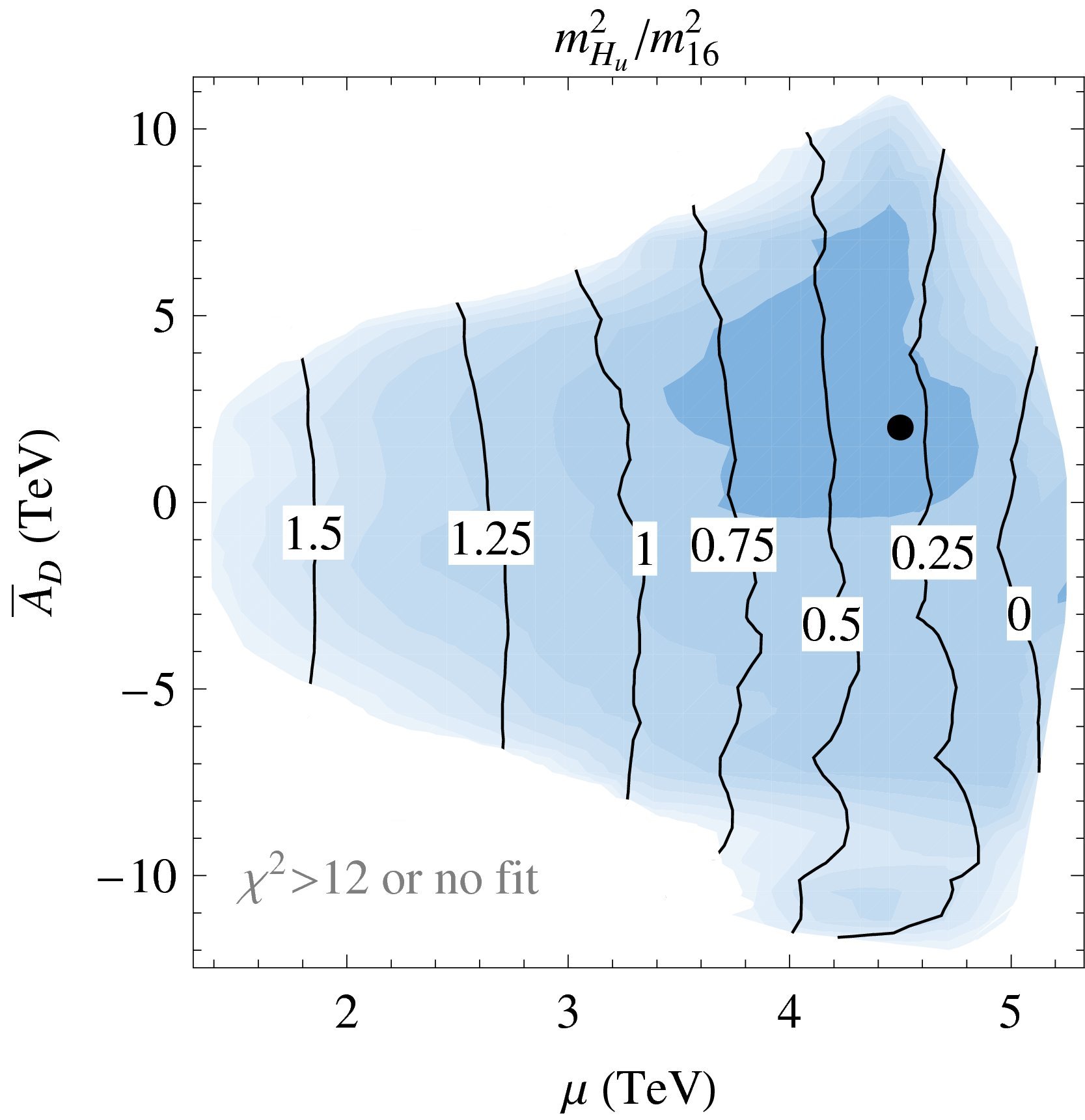}
\hspace{0.04\textwidth}
\includegraphics[width=0.45\textwidth]{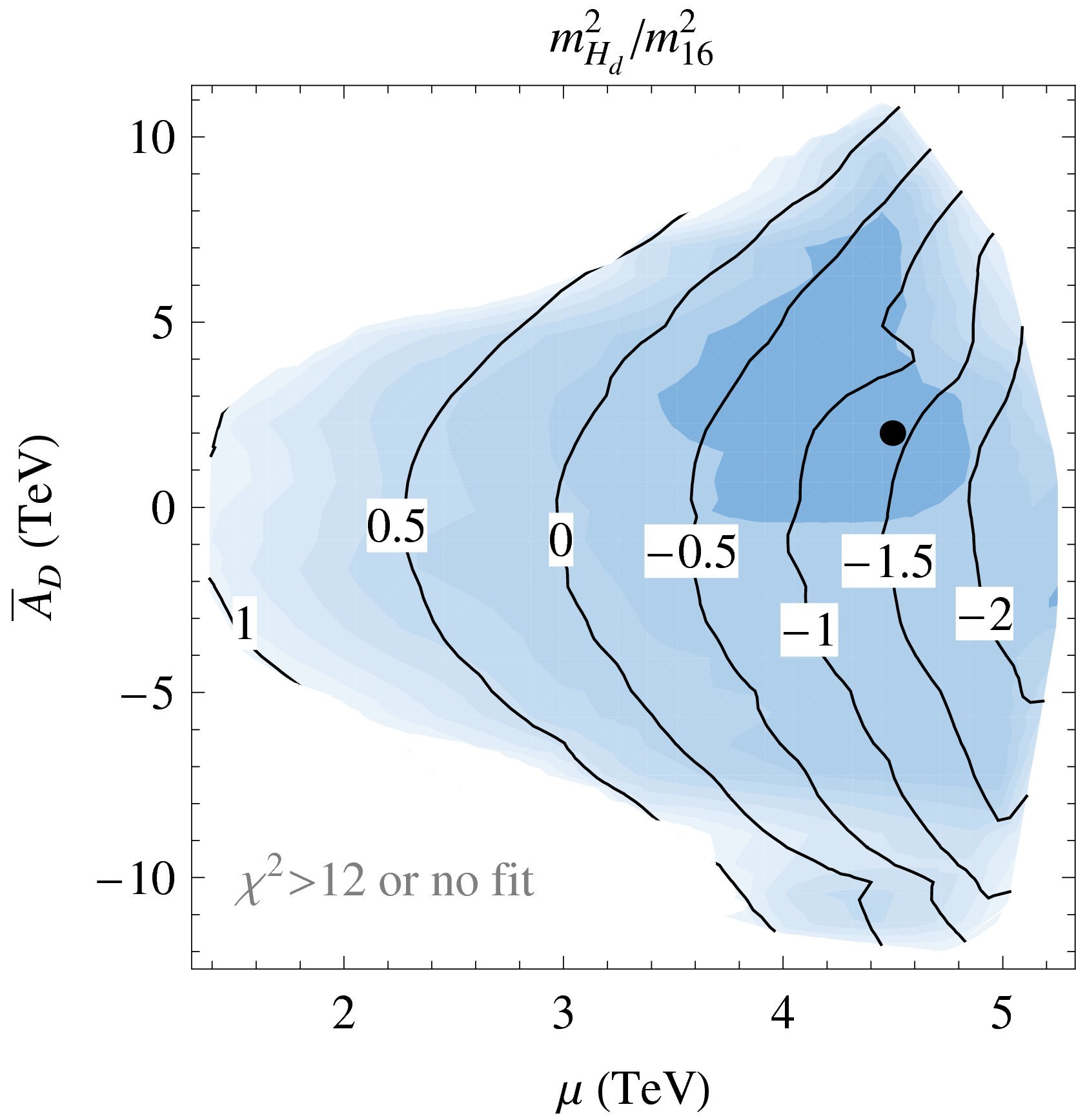}
\caption{Lines of constant values for four of the input parameters, superimposed on the total 
$\chi^2$ map (cf. figure~\ref{fig:contours-chi2}). Top left: $\ov A_U$ in TeV. Top right: $m_{1/2}$ 
in GeV. Bottom left: $m_{H_u}^2/m_{16}^2$. Bottom right: $m_{H_d}^2/m_{16}^2$.}
\label{fig:contours-inp}
}

\FIGURE[ht]{
\includegraphics[width=0.45\textwidth]{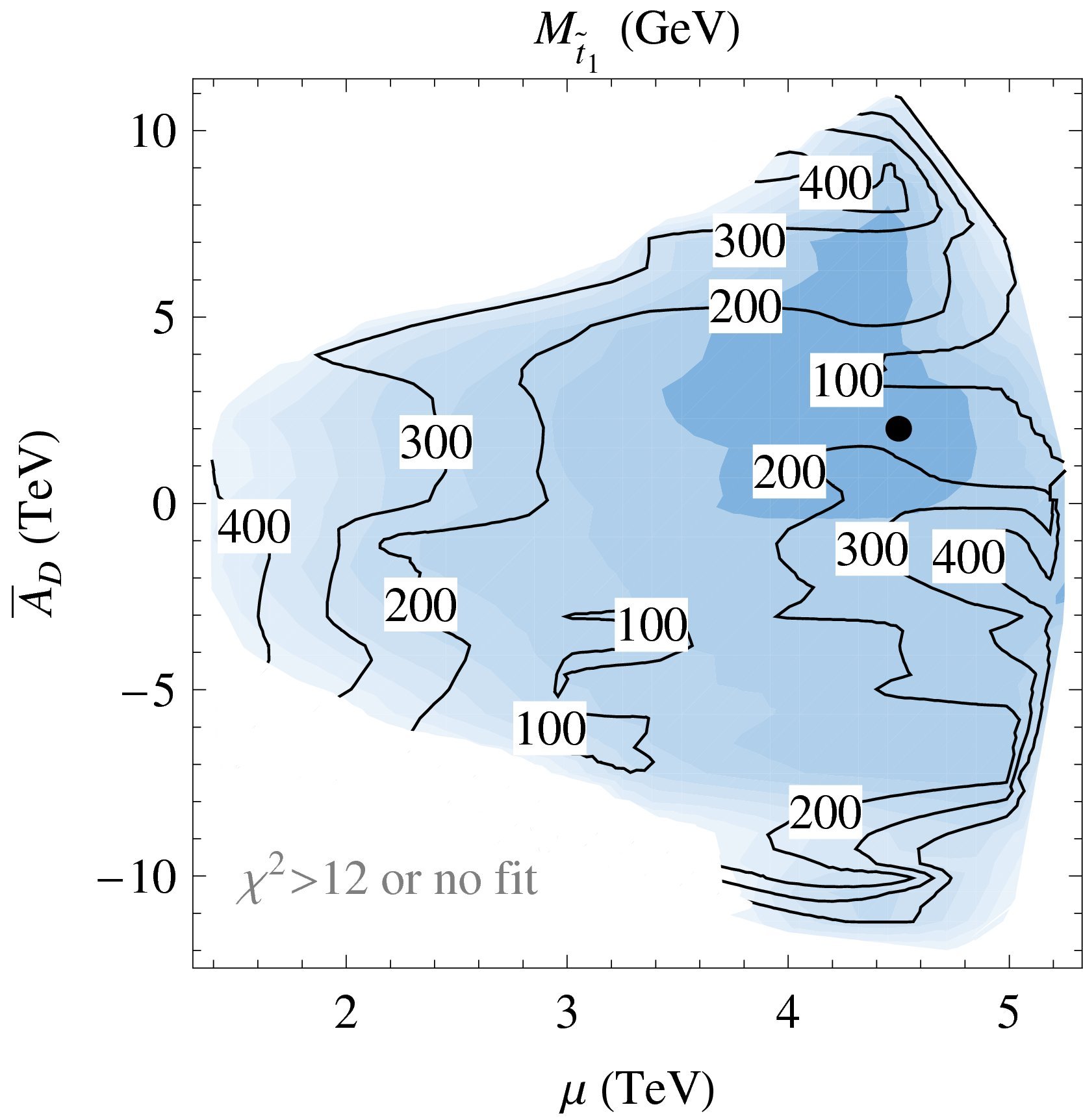}
\hspace{0.04\textwidth}
\includegraphics[width=0.45\textwidth]{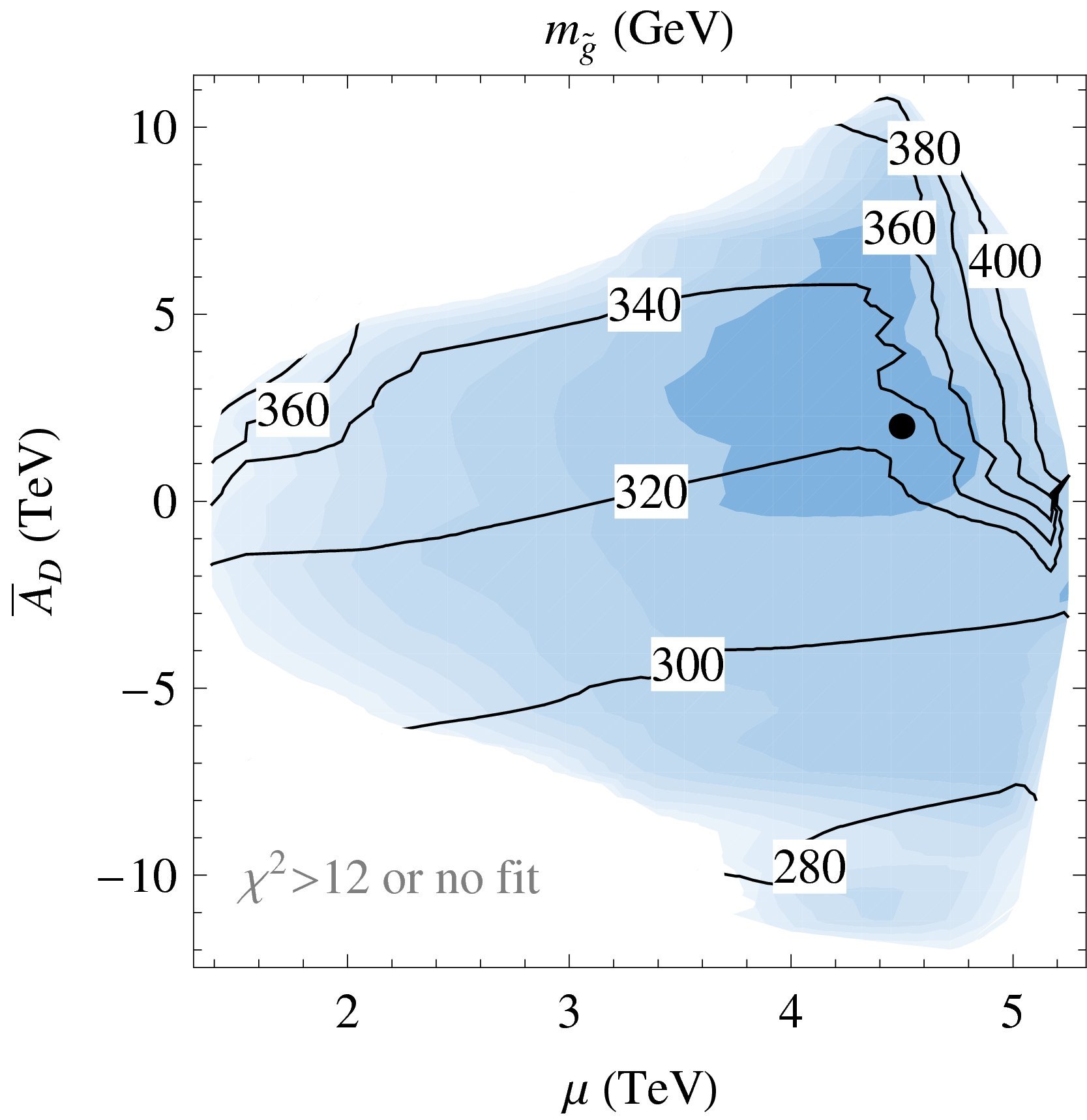}
\caption{Lines of constant masses (in GeV) chosen by the fit for the lightest stop (left) and the gluino (right).}
\label{fig:contours-obs}
}

\subsection{Main results}\label{sec:mainresults}

The left-hand panel of fig. \ref{fig:contours-chi2} displays the lines of constant $\chi^2$ 
(as defined in eq. (\ref{eq:chi2})) in the $\mu$ vs $\ov A_D$ plane. This plot is obtained by 
sampling the $\chi^2$ function on this plane via MIGRAD minimizations where only $m_{16}$, 
$\mu$ and $\ov A_D$ are kept fixed. The right-hand panel of fig. \ref{fig:contours-chi2} shows 
the contribution to the $\chi^2$ function solely from the BR($B \to X_s \gamma$) constraint, 
backdropped by the total $\chi^2$ contours for comparison. In fig. \ref{fig:contours-inp}, 
we show the contours of the four most important input parameters of table~\ref{tab:parameters} 
chosen by the fitting procedure. Finally, in figures \ref{fig:contours-obs}--\ref{fig:contours-obs2}, 
we show the contours of several masses of interest.

Let us make some immediate observations.
\begin{enumerate}

\item\label{it:1} In the left-hand panel of fig. \ref{fig:contours-chi2}, the lowest values for 
the $\chi^2$ function are obtained for $\mu \approx 3.5-4.5$ TeV, which is of the order of $m_{16}$, 
here set to 4 TeV.

\item\label{it:2} The $\chi^2$ contours are roughly symmetric with respect to the axis $\ov A_D=0$.

\item\label{it:3} From the top-left panel of fig. \ref{fig:contours-inp}, one can see that the 
preferred region points to $\ov A_U \approx -2.5 \, m_{16}$, and in particular to a sizable $\ov A_U-\ov A_D$ 
splitting. The bad phenomenological performance of YU in the case of universality between 
trilinears \cite{AABuGuS,AlGuRaS} is recovered in the limit $\ov A_D \to \ov A_U \approx - 2 \, m_{16}$, 
and $\mu \ll m_{16}$, i.e. in the lower, leftmost part of the plot.

\item\label{it:4} From the right-hand panel of fig. \ref{fig:contours-chi2}, one can see that the 
BR($B \to X_s \gamma$) constraint is the main driving force for large $\mu$. Note that, with our assumptions 
on the theoretical uncertainties as in table~\ref{tab:obs-FC}, $\chi^2_{b\to s \gamma}=1$ (2) corresponds to 
$\text{BR}(B \to X_s \gamma) \simeq 3.0 \times 10^{-4}$ ($2.5 \times 10^{-4}$). The branching ratio is always 
below the experimental central value due to the destructive interference between SM and chargino contributions.

\item\label{it:5} The top-right panel of fig. \ref{fig:contours-inp} shows that the universal 
gaugino mass $m_{1/2}$ increases with increasing $\ov A_D$, and is also dependent on the sign 
of $\ov A_D$. For positive $\ov A_D$ and $\mu>4$ TeV, there is a steep rise in $m_{1/2}$.

\item\label{it:6} For $\mu \gtrsim 4.5$ TeV, the $\chi^2$ starts deteriorating again, although 
the $\chi^2$ contribution from BR($B \to X_s \gamma$) is very small in this region. Both the stop 
mass and the gluino mass, as shown respectively in the left and right panels of fig. \ref{fig:contours-obs}, 
increase again in this region.

\item\label{it:7} The two plots at the bottom of fig. \ref{fig:contours-inp} show that $m_{H_u}^2$ 
is preferred to be positive and equal to or less than $m_{16}^2$, while $m_{H_d}^2$ is preferred to 
be zero or even negative at the GUT scale. This is in contrast to refs. \cite{AABuGuS,AlGuRaS}, 
where both $m_{H_u}^2$ and $m_{H_d}^2$ were positive and greater than $m_{16}^2$. $m_{H_u}^2$ 
is basically independent of $\ov A_D$.

\end{enumerate}

Points \ref{it:1}.--\ref{it:3}. above confirm the previously obtained results \cite{AlGuRaS} 
that YU is disfavoured for $\ov A_U=\ov A_D$ by an interplay between the corrections to the 
bottom quark mass and the FCNC constraints, but they also show that the trilinear splitting 
scenario considered here indeed gives rise to a viable solution featuring exact YU and being 
compatible with all relevant constraints. Interestingly enough, the recovery of phenomenological 
viability is not obtained by invoking a decoupling of the sparticle spectrum, but it instead 
seems to {\em require} parts of this spectrum to be very close to their experimental lower bounds.

\TABLE[ht]{
\begin{tabular}{|llll|}
\hline
Observable  &  Exp.  &  Fit  &  Pull  \\
\hline\hline
$M_W$  &  80.398  &  80.58  &  0.5  \\
$M_Z$  &  91.1876  &  90.65  &  \textbf{1.2}  \\
$10^{5}\; G_\mu$  &  1.16637  &  1.164  &  0.4  \\
$1/\alpha_\text{em}$  &  137.036  &  136.7  &  0.5  \\
$\alpha_s(M_Z)$  &  0.1176  &  0.1176  &  0.0  \\
$M_t$  &  173.1  &  172.7  &  0.3  \\
$m_b(m_b)$  &  4.20  &  4.22  &  0.3  \\
$M_\tau$  &  1.777  &  1.78  &  0.1  \\
$10^{4}\; \text{BR} (B \to X_s \gamma)$  &  3.52  &  3.04  &  0.9  \\
$10^{6}\; \text{BR} (B \to X_s \ell^+\ell^-)$  &  1.60  &  1.63  &  0.0  \\
$\Delta M_s / \Delta M_d$  &  35.1  &  33.9  &  0.3  \\
$10^{4}\; \text{BR} (B^+ \to \tau^+\nu)$  &  1.40  &  0.93  &  \textbf{1.0}  \\
$10^{8}\; \text{BR} (B_s \to \mu^+\mu^-)$  &  $<5.8 $ &  2.01  &  --  \\
\hline
\multicolumn{3}{|r}{total $\chi^2$:}  &  \textbf{4.05} \\
\hline
\end{tabular}
\caption{Example of fit in the region with successful YU. The pull in the last column is defined as the square root of the $\chi^2$ contribution.}
\label{tab:examplefit}
}

\TABLE[ht]{
\renewcommand\arraystretch{1.05}
\begin{tabular}{|lc|lc|}
\hline
\multicolumn{2}{|l}{Input parameters} & \multicolumn{2}{|l|}{Spectrum predictions} \\
\hline
\hline
$m_{16}$  & 4000 & $M_{h^0}$  &  126  \\
$\mu$  & 4500 & $M_{H^0}$  &  1109  \\
$m_{1/2}$  &113.8& $M_{A}$  &  1114  \\
$\ov A_D$   &2000& $M_{H^+}$  &  1115  \\
 $\ov A_U$ & $-11321$ & $M_{\tilde t_1}$  &  192  \\
$\tan\beta$  &49.8& $m_{\tilde t_2}$  &  2656  \\
$1/\alpha_G$  &24.7& $m_{\tilde b_1}$  &  2634  \\
$M_G / 10^{16}$  &3.77& $m_{\tilde \tau_1}$  &  3489  \\
$\epsilon_3 / \%$  &$-3.8$& $m_{\tilde\chi^0_1}$  &  53.3  \\
$(m_{H_u}/m_{16})^2$  &$0.32$& $m_{\tilde\chi^0_2}$  &  104.1  \\
 $(m_{H_d}/m_{16})^2$  &$-1.38$& $m_{\tilde\chi^+_1}$  &  104.1  \\
$y_t$  &0.66& $m_{\tilde g}$  &  321  \\
$M_{R} / 10^{14}$  & $2.6$&  & \\
\hline
\end{tabular}
\caption{Input parameters and spectrum predictions for the example fit reported in table~\ref{tab:examplefit}. 
All masses and massive input parameters are in units of GeV.}
\label{tab:examplefit-par}
}

In table~\ref{tab:examplefit}, we report the fitted values for the observables entering 
the $\chi^2$ function for one fit belonging to the region with lowest $\chi^2$. The input values for this example fit
are reported in the left panel of table~\ref{tab:examplefit-par} and the resulting spectrum predictions, on which
we will comment again in section \ref{sec:spectrum}, on the right panel of the same table. This example fit is 
also represented in figures \ref{fig:contours-chi2}--\ref{fig:contours-obs} as a dot.
Apart from a pull in the $Z$ mass, the largest contributions to the $\chi^2$ come from $B\to X_s\gamma$ 
and $B\to\tau\nu$. However, it should be noted that in both of these cases, the SUSY contribution at this 
parameter point is very small and most of the `discrepancies' are rooted in discrepancies between the 
current experimental central values and the SM prediction. The prediction for $B_s\to\mu^+\mu^-$ is well below 
the current experimental upper bound (cf. table~\ref{tab:obs-FC}) but well within the LHCb (and probably 
even Tevatron) reach. Since the predictions are in the $(1-3)\times 10^{-8}$ ballpark in the entire 
preferred region, an experimental upper bound below $10^{-8}$ would put this scenario in question.

In the following section, we will attempt to interpret the above findings on theoretical grounds.

\subsection{Interpretation of fit results}\label{sec:interpretation}

First of all, let us discuss the items \ref{it:1}.--\ref{it:3}. of the above observations, i.e. 
the questions why the trilinear splitting helps to obtain successful YU, why this mechanism is 
roughly symmetric under a sign change of $\ov A_D$ and why it requires large $\mu$. The
mechanism at work can be understood by recalling the basic ingredients for YU discussed in 
section~\ref{sec:YU}, and it turns out to be quite compelling.

Recall that, to suppress the gluino corrections to $m_b$ with respect to the chargino ones, 
a hierarchy $m_{\tilde t_i}\ll m_{\tilde b_i}$ is required in addition to a large trilinear 
parameter for the stop at the electroweak scale. This hierarchy is dependent on the 
trilinear parameters of the up- and down-type squarks, because the latter contribute to the 
RG evolution of the running squark masses. Assuming $\tan\beta\approx 50$ and neglecting gaugino 
mass contributions (anticipating the condition $m_{1/2}\ll m_{16}, \ov A_{U,D}$), the low-energy 
values for the third-generation squark mass-squared parameters can be approximately written in 
terms of the GUT-scale parameters as
\beqn
\label{eq:mQ2-approx}
\left(m_{Q}^2 \right)_{33} &\approx& 0.51 \, m_{16}^2 -0.12 \, m_{H_u}^2 -0.09 \, m_{H_d}^2 -0.02 \, \ov A_U^2 -0.02 \, \ov A_D^2~, \\
\label{eq:mU2-approx} 
\left(m_{U}^2 \right)_{33} &\approx& 0.49 \, m_{16}^2 -0.22 \, m_{H_u}^2 -0.01 \, m_{H_d}^2 -0.06 \, \ov A_U^2 +0.01 \, \ov A_D^2~,\\
\label{eq:mD2-approx}
\left(m_{D}^2 \right)_{33} &\approx& 0.55 \, m_{16}^2 +0.01 \, m_{H_u}^2 -0.21 \, m_{H_d}^2 +0.01 \, \ov A_U^2 -0.05 \, \ov A_D^2~.
\eeqn
As is apparent from eq. (\ref{eq:mU2-approx}), a very light right-handed stop can be obtained by appropriately adjusting 
$m_{H_u}^2$ and $\ov A_U^2$ at the GUT scale. In the universal case, $\ov A_U^2 = \ov A_D^2 \equiv A_0^2$, a sizable $A_0$ 
also leads to a reduction of $(m_{D}^2)_{33}$ and $(m_{Q}^2)_{33}$. Instead, in the non-universal case, the choice
$\ov A_D^2 \ll \ov A_U^2$ allows to maintain a light right-handed stop, while preventing negative RGE contributions to the 
right-handed sbottom and left-handed squark masses. As a result, this mechanism permits to obtain a strong mass hierarchy
\beq
\left(m_{U}^2 \right)_{33} \ll \left(m_{Q}^2 \right)_{33} < \left(m_{D}^2 \right)_{33},
\eeq
implying\footnote{Note that the LR mixing terms in the squark mass matrices do not play a role in this discussion, 
since $m_{16}^2 \gg m_t A_t$ and $\gg m_b \mu \tan\beta$.}
\beq
m_{\tilde t_R} \ll m_{\tilde t_L} \approx  m_{\tilde b_L} <  m_{\tilde b_R},
\label{eq:mtbLR}
\eeq
which is what is \textit{needed} to maximize the negative chargino corrections to $m_b$ and suppress the gluino contributions. 
While this hierarchy is also present in the universal case, it can be greatly amplified in the trilinear splitting scenario 
by reducing $\ov A_D^2$: this leads to only a mild increase of $ m_{\tilde t_L}$ and  $m_{\tilde b_L}$, but a strong increase 
of $m_{\tilde b_R}$, while leaving  $m_{\tilde t_R}$ almost unaffected.

On the basis of equations (\ref{eq:mQ2-approx})--(\ref{eq:mD2-approx}), this RG effect is manifestly invariant under a 
sign change in either $A_U$ or $A_D$. This invariance -- which explains the approximate symmetry observed in item \ref{it:2}. --
is in fact due to the condition $m_{1/2}\ll \ov A_{U,D}$, which is preferred by Yukawa unification, and would otherwise be 
spoiled by terms proportional to $m_{1/2}\ov A_{U,D}$ in equations (\ref{eq:mQ2-approx})--(\ref{eq:mD2-approx}).

We stress at this point that, in our numerical analysis, we calculate the physical mass of the light stop at the one-loop level, 
similarly to our procedure in ref. \cite{AlGuRaS}, since the lightness of the fitted stop mass implies that one-loop 
corrections are crucial to assess whether a given parameter point is viable or excluded by a tachyonic stop. For the remaining 
sparticles, we use the running masses.

Once the above mechanism has ensured that the overall sign of the threshold correction to $m_b$ is {\em negative}, an increase 
in $\mu$, to which the threshold correction is proportional (see eqs.  (\ref{eq:deltamb-g}) and (\ref{eq:deltamb-ch})) helps 
to make it parametrically {\em large enough} in magnitude, as required to fit the experimental data. At the same time, and quite
interestingly, this large $\mu$ suppresses the chargino contributions to the $b\to s\gamma$ amplitude (see \cite{WA} for a discussion
on this point), therefore preventing a large destructive interference with the SM contribution.\footnote{For large $\mu \times 
\tan \beta$ one may worry about the size of gluino contributions as well \cite{DudleyKolda,WA}. In our case, gluino contributions 
have roughly the same size as those from Higgses (in turn, an order of magnitude smaller than charginos), but opposite sign, thus 
cancelling with each other almost exactly.} This makes clear at the same time why the BR$(B\to X_s\gamma)$ constraint dominates 
the $\chi^2$ for small $\mu$, as observed in point \ref{it:4}. of section~\ref{sec:mainresults} and shown in the right-hand panel 
of fig. \ref{fig:contours-chi2}. We stress that, similarly to ref. \cite{AlGuRaS}, we allow for both signs of the 
$b\to s\gamma$ amplitude in our analysis. The large $\mu$ solution described above is preferred by the fit over the solution 
where the SUSY contributions are so large that they flip the sign of the $b\to s\gamma$ amplitude. In fact, the latter case turns 
out to imply a too large branching ratio of $B\to X_s\ell^+\ell^-$ with respect to the experimental measurement \cite{Gambino:2004mv}.

Concerning point \ref{it:5}. of section~\ref{sec:mainresults}, the value of $m_{1/2}$ chosen by the fit, this is due to the 
requirement of a light gluino mass to suppress the gluino corrections to $m_b$, as discussed in section~\ref{sec:YU}.
In fact, $m_{1/2}$ is always fitted close to its lowest allowed value, set by the LEP lower bound on the mass of the lightest 
chargino (see table \ref{tab:obs-EW}), which is an almost pure Wino in our setup, due to the large $\mu$. If only one-loop 
RGEs for the gaugino masses were used, $M_2 > 104$ GeV would imply $m_{1/2} \gtrsim 132$ GeV. However, due to the conditions 
$|\ov A_{U,D}| \gg m_{1/2}$, two-loop effects become important in the running of the gaugino masses. These two-loop 
contributions are responsible both for the possibility of having $m_{1/2}$ less than 132 GeV and for the rise of $m_{1/2}$ with 
$\ov A_D$, visible in the top-right panel of fig. \ref{fig:contours-inp}. This effect always ensures a light chargino, except
in the top-right corner of the plot. This region will be discussed in the next paragraph.

For too large $\mu$, the $\chi^2$ starts worsening again, as mentioned in point \ref{it:6}. of section~\ref{sec:mainresults}: 
in fact, in this region, the negative corrections to $m_b$ start being so large that the mechanism above has to be tamed 
to prevent $m_b$ from dropping below 4.2~GeV. There are different possibilities to achieve this: for $\mu>4.5$ TeV and $\ov A_D<0$, 
the fits tune the lightest stop mass to be larger, as shown in the left panel of fig. \ref{fig:contours-obs}, reducing the 
size of the chargino corrections to $m_b$; for $\mu>4.5$ TeV and $\ov A_D>0$, the fits instead increase $m_{1/2}$ and accordingly 
the gluino mass, as shown in the right panel of fig. \ref{fig:contours-obs}, increasing the gluino corrections to $m_b$. 
While these mechanisms allow to obtain a correct value for $m_b$, they cause tensions in other observables, leading to a steep 
rise in the $\chi^2$. Therefore, viable YU solutions at even higher $\mu$ are not to be expected.

Regarding the values of $m_{H_{u,d}}^2$ commented on in point \ref{it:7}. of section~\ref{sec:mainresults}, $m_{H_u}^2$ is 
basically independent of $\ov A_D$ because its value is fixed by the EWSB conditions. Indeed, at large $\tan\beta$, these 
conditions require $m_{H_u}^2 \approx - |\mu|^2$ to hold at the EW scale.
The value of $m_{H_d}^2$ on the other hand is bounded from below because EWSB requires 
$m_{H_d}^2 \gtrsim m_{H_u}^2$ at the EW scale and bounded from above because a too large value would drive the sbottom 
masses smaller, cf. eq.~(\ref{eq:mD2-approx}), which is unfavourable for YU.

The possibility to indeed fulfill the weak-scale conditions $-|\mu|^2 \approx m_{H_u}^2 \lesssim m_{H_d}^2$, thereby achieving 
correct EWSB, can be illustrated through the following approximate expressions for the Higgs soft terms
\beqn
\label{mHu-approx}
m_{H_u}^2(M_Z) &=&
  -0.74 \, m_{16}^2  +0.56 \, m_{H_u}^2  + 0.06 \, m_{H_d}^2
  -0.11 \, \overline A_U^2  + 0.01 \, \overline A_D^2~, \\
\label{mHd-approx}
m_{H_d}^2(M_Z) &=&
  -0.81 \, m_{16}^2  +0.06 \, m_{H_u}^2  + 0.52 \, m_{H_d}^2  
  +0.01 \, \overline A_U^2  -0.14 \, \overline A_D^2~,
\eeqn
where the parameters on the r.h.s. of either equation are at the GUT scale. These expressions reproduce with remarkable accuracy
the Higgs soft terms calculated with the full numerical procedure and may be used, e.g., on the parameter values of the example 
fit in table \ref{tab:examplefit-par}. 
\FIGURE[ht]{
\includegraphics[width=0.45\textwidth]{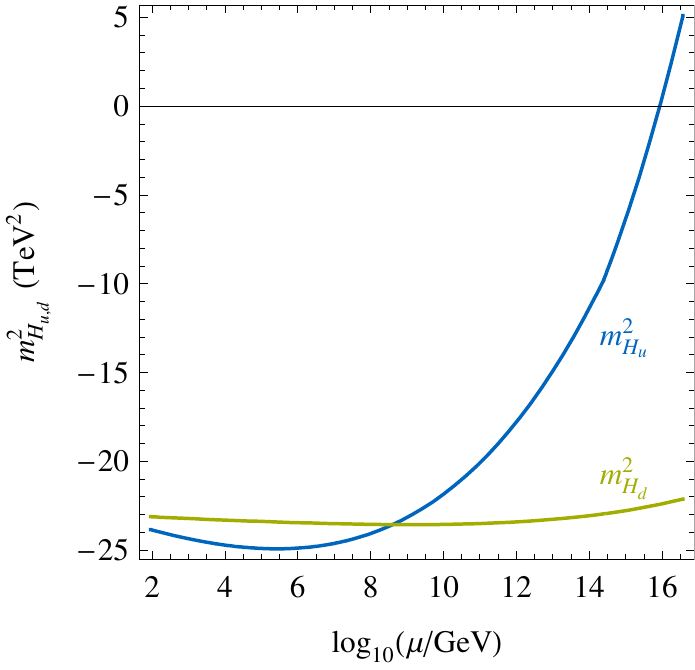}
\caption{RG evolution of the parameters $m_{H_{u,d}}^2$ for the example fit in tables \ref{tab:examplefit}-\ref{tab:examplefit-par}.}
\label{fig:mHudRGE}
}
For this fit, the evolution of the Higgs soft terms to low scales is also reported in figure \ref{fig:mHudRGE}. The latter shows
in particular that $m_{H_d}^2$ stays approximately constant throughout the running range, at variance with $m_{H_u}^2$, thus eventually 
allowing the weak-scale inequality $m_{H_u}^2 \lesssim m_{H_d}^2$. Hence from this figure and table \ref{tab:examplefit} 
one can conclude that EWSB can indeed be quantitatively fulfilled throughout the explored parameter space, with pulls on EW observables 
approximately constant in the best-fit region. \footnote{We note that inclusion of the one-loop tadpoles \cite{PBMZ} in 
the EWSB equations plays in our case an important role, since these tadpoles will receive large logarithmic corrections from various 
among the heavy particles present in the spectrum. We thank the Referee for triggering this discussion.}

Considering that the natural EW scale is sensibly below the $m_{16}$- and $\mu$-ranges preferred by the mechanism discussed above, 
it is clear that our scenario involves some degree of fine tuning in order for EWSB to be successful at the quantitative
level.\footnote{We observe, however, that a certain amount of fine tuning in EWSB is by now a common feature within low-energy 
SUSY (for an insightful discussion, see \cite{GR}).} We also note that, in the case of the trilinear splitting scenario, the 
inverse scalar mass hierarchy, i.e. the hierarchy between the third and the first two generations sfermions, is reduced because 
of the heavier masses for the sbottoms and staus. This hierarchy, which improves the fine-tuning in the corrections to 
the Higgs mass, would instead be active in the case of universal trilinear couplings \cite{BDR1,BDR2,AABuGuS,AlGuRaS}.
We emphasize, on the other hand, that the trilinear splitting mechanism implies a very light stop
mass, with a Higgs mass comfortably above the LEP bound, and with all FCNC constraints automatically fulfilled. On a model-dependent
basis, all these desired features would require corresponding amounts of fine tuning as well, that in our case are simply absent.

To summarize, the mechanisms described above single out a region in parameter space where successful YU is obtained, standing
all the other experimental constraints, by a non-trivial interplay between the requirements of a large enough negative correction 
to $m_b$ and a small enough correction to $b\to s\gamma$. As we saw above, allowing for split trilinear couplings, this can be 
achieved by $|\ov A_D|^2 \ll |\ov A_U|^2$ {\em and} large $\mu \approx m_{16}$ (but not too large). We emphasize that the recovery of 
phenomenological viability is not obtained by invoking a decoupling of the sparticle spectrum, it instead strongly {\em requires} 
parts of this spectrum to be very close to their experimental lower bounds. Since this observation is crucial for the LHC phenomenology 
of this class of models, we now briefly discuss the SUSY spectrum entailed by successful YU.

\FIGURE[ht]{
\includegraphics[width=0.45\textwidth]{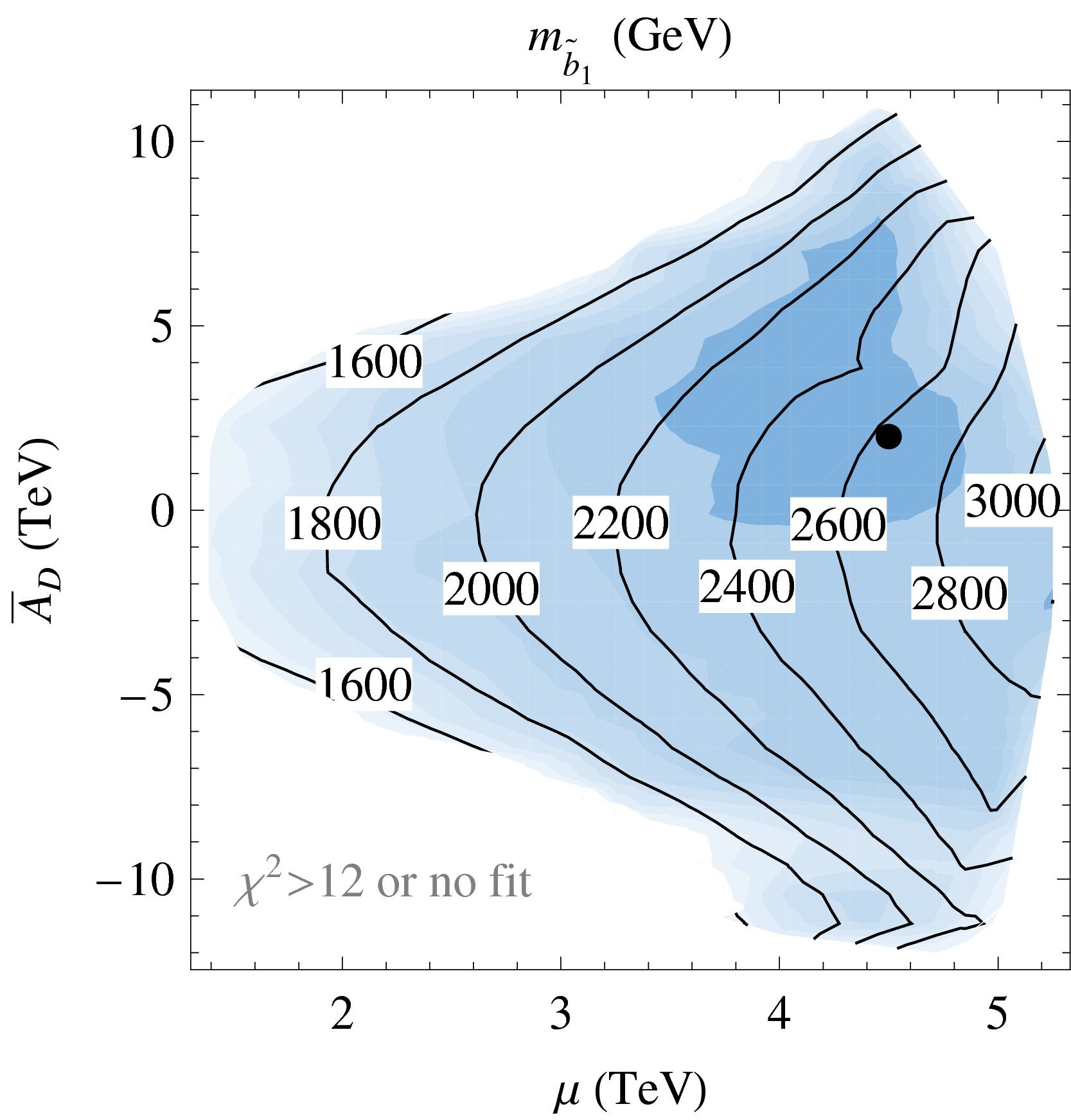}
\hspace{0.04\textwidth}
\includegraphics[width=0.45\textwidth]{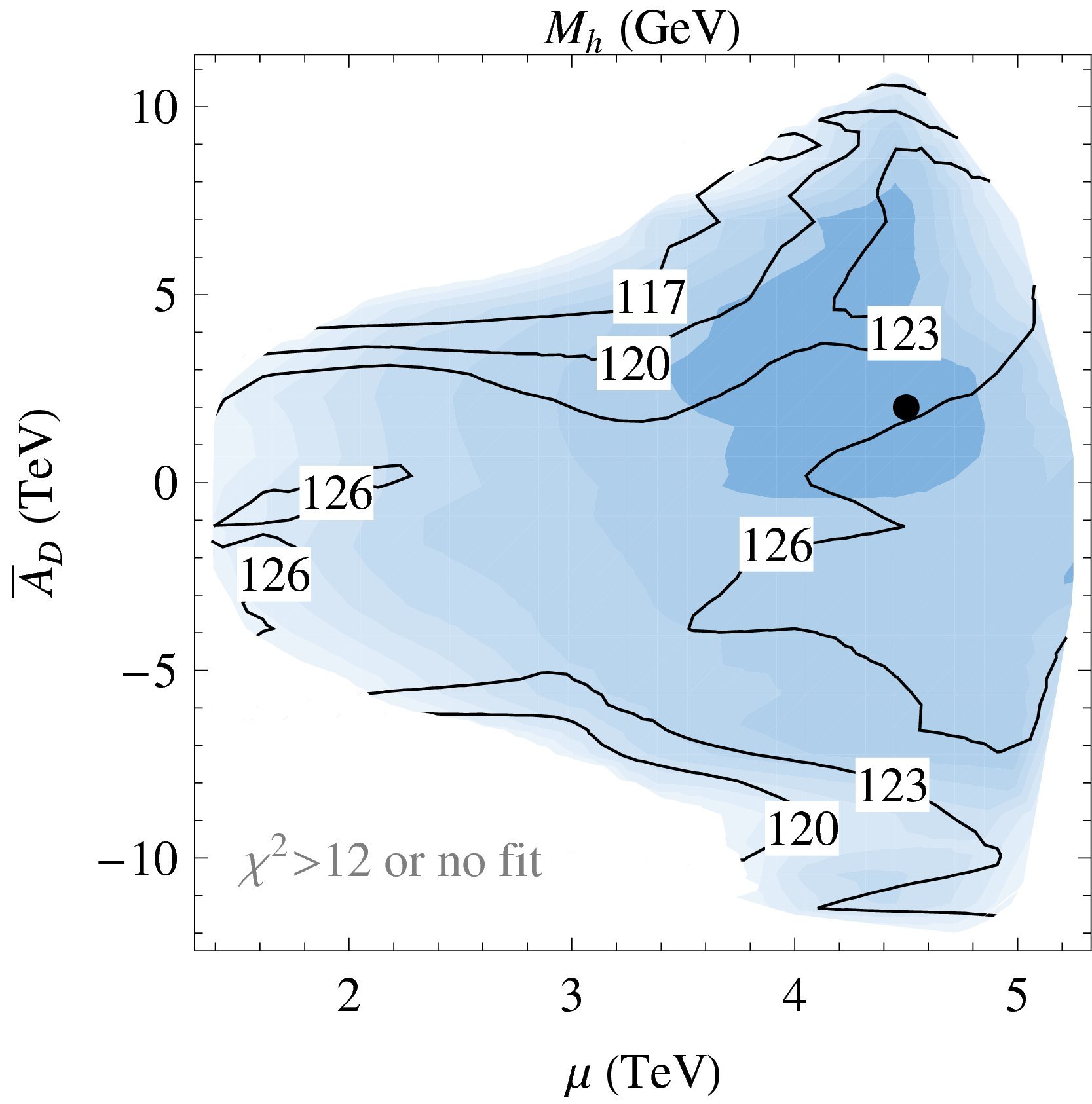}
\caption{Lines of constant masses (in GeV) chosen by the fit for the lightest sbottom (left) and the light neutral Higgs (right).}
\label{fig:contours-obs2}
}

\subsection{Sparticle spectrum with successful YU}\label{sec:spectrum}

As discussed in the previous sections, the region in parameter space favouring YU features a gluino with a mass of around 350~GeV 
and a very light stop. In fact, as explained, a light stop mass is one of the main ingredients of the mechanism that allows successful
YU in our trilinear splitting scenario. Therefore, a stop mass as low as about 100--200~GeV is generally preferred. However, as can bee seen 
in the left-hand panel of fig.~\ref{fig:contours-obs}, more generally stop masses $\lesssim$ 400~GeV can lead to viable Yukawa unification. 
The large stop-sbottom splitting required by YU, as discussed in 
section~\ref{sec:YU}, can be appreciated in fig.~\ref{fig:contours-obs2}, showing the contours of the lightest sbottom mass, which is always 
nearly degenerate with the heavy stop, cf. (\ref{eq:mtbLR}). The lightest chargino and light neutralino masses are preferred to be as light 
as is allowed by experiment. In the right-hand panel of fig.~\ref{fig:contours-obs2}, we show the contours of the light neutral Higgs boson mass. 
Interestingly enough, it is found to be in the 120--125~GeV region, such that the LEP constraint is not active. The heavy neutral, charged 
and pseudoscalar Higgs bosons are nearly degenerate and assume values around 1~TeV. All these predictions are exemplified in the right
panel of table \ref{tab:examplefit-par}. Concerning the part of the SUSY spectrum not reported in this table, masses lie in the ballpark of
$m_{16}$: for first and second generation sfermions, because of the absence of a large Yukawa contribution in the running;
for the heavy neutralinos as well as the heavy chargino, because they are mostly Higgsinos, with masses of O($\mu$).

The non-decoupling and clear-cut nature of these spectrum predictions make the split trilinear scenario a compelling and testable one.
In this respect, a first interesting question is whether part of our stop vs gluino parameter space may actually not comply with the SUSY mass bounds
provided by Tevatron (see e.g. \cite{Aaltonen:2008rv}), which are stronger than those in our table \ref{tab:obs-EW}. We note however that the 
Tevatron bounds typically assume mSUGRA scenarios and we cannot identify any obvious relation to translate those bounds to our case. We 
believe that this issue would deserve a separate study.

\subsection{The role of right-handed neutrinos}\label{sec:RHnu}

At this point, it is worth commenting on the effect of our inclusion of right-handed neutrinos, as discussed in section~\ref{sec:procedure}, 
on the numerical results.

Neutrino Yukawa couplings enter in the RGEs of up-type quark and charged lepton Yukawas and tend to drive these couplings to smaller values. 
In the leading-log approximation, the difference between the values of the top and tau Yukawa couplings at low energies in the presence of 
neutrino Yukawa contributions and the values they would take in the absence of right-handed neutrinos is
\beq
y_t - y_t^{0\nu} = y_\tau - y_\tau^{0\nu} =  -\frac{1}{16\pi^2} \, y_t(M_G) \, \log \left( \frac{M_G}{M_R} \right) \,.
\eeq
This percent level change in Yukawa couplings can be compensated by adjusting accordingly the GUT-scale value of the Yukawa coupling and 
$\tan\beta$, which in turn affects the value of the $b$ quark Yukawa coupling $y_b$ at low energies.

So, while right-handed neutrinos in principle affect the evolution of the third generation Yukawa couplings, we found numerically that the 
presence of the neutrino Yukawa contributions and the value of the right-handed neutrino mass scale $M_R$ do not significantly affect the 
success of YU for any given parameter point, since the small changes in Yukawa couplings induced by these contributions can easily be 
compensated by changes in the remaining input parameters.

In the soft sector, the neutrino Yukawa contributions leave the largest impact on the left-handed slepton doublet mass term $m_L^2$ and 
on the up-type Higgs mass term $m_{H_u}^2$. In the leading-log approximation and with the boundary conditions (\ref{eq:ST-m2})--(\ref{eq:ST-A}), 
their low-energy values are modified according to \cite{Hisano95,Petcov03}
\beq
\left( m_L^2 \right)_{ij} - \left( m_L^2 \right)_{ij}^{0\nu} = 
-\frac{1}{16\pi^2}\left( 4 m_{16}^2 + 2 m_{H_u}^2 + 4 \ov A_U^2 \right) ( Y_\nu^\dagger Y_\nu )_{ij}  \log \left( \frac{M_G}{M_R} \right) \,,
\label{eq:Ynu-mL2}
\eeq
\beq
m_{H_u}^2 - (m_{H_u}^2)^{0\nu} = 
-\frac{1}{16\pi^2}\left( 4 m_{16}^2 + 2 m_{H_u}^2 + 4 \ov A_U^2 \right) \text{Tr}( Y_\nu^\dagger Y_\nu )  \log \left( \frac{M_G}{M_R} \right) \,,
\label{eq:Ynu-mHu2}
\eeq
where the quantities on the right-hand side are defined at the GUT scale. According to (\ref{eq:Ynu-mL2}), the presence of right-handed neutrinos 
leads to lighter left-handed sleptons at low energies; however, this does not have any relevant impact on the mechanism ensuring the success of YU. 
The off-diagonal components of (\ref{eq:Ynu-mL2}) give rise to lepton flavor violating decays \cite{Hisano95}, but definite predictions 
can only be made in models predicting $( Y_\nu^\dagger Y_\nu )_{ij}$.
Eq. (\ref{eq:Ynu-mHu2}) shows that the neutrino Yukawa contributions drive $m_{H_u}^2$ to smaller values. However, this can be easily compensated 
by raising the value of $m_{H_u}^2$ at the GUT scale, which is possible in the setup of non-universal Higgs masses. We stress that this change in 
$m_{H_u}^2$ induced by right-handed neutrino effects is not sufficient to explain or to generate the large $m_{H_u}^2-m_{H_d}^2$ splitting 
required for successful YU.

To summarize, our approach of taking into account contributions from right-handed neutrinos on the evolution of couplings introduces one 
more free parameter, $M_R$, which allows to account for neutrino induced threshold corrections in particular to Yukawa couplings and to 
$m_{H_u}^2$. However, numerically, the fits turn out to be quite insensitive to the value of this parameter and even a removal of 
right-handed neutrino effects by taking $M_R\to M_G$ does not significantly affect the results. This means on the one hand that YU does 
not prefer or single out a particular scale for right-handed neutrinos; on the other hand, it means that the mechanism identified in our 
analysis cannot be spoiled by right-handed neutrino effects.

\subsection[Remarks on $(g-2)_\mu$ and dark matter relic abundance]{\boldmath Remarks on $(g-2)_\mu$ and dark matter relic abundance}\label{sec:gm2}

Before concluding this section, we would like to comment on two observables we did not take into account in our numerical analysis: the muon 
anomalous magnetic moment $a_\mu = (g-2)_\mu$ and the neutralino cosmological relic density.

Under our assumptions of sfermion mass universality, cf. (\ref{eq:ST-m2}), we found a posteriori predictions for the SUSY contributions to 
$a_\mu$ at the $10^{-11}$ level, much smaller than the current, O($10^{-9}$) discrepancy between experiment and the SM 
prediction \cite{Jegerlehner:2009ry}. The reasons for this smallness are on the one hand the large values of the $\mu$ parameter and on the 
other the relatively heavy slepton spectrum, both leading to a suppression of the chargino contributions to $a_\mu$. We note however that a 
relaxation of the universality assumption (\ref{eq:ST-m2}) and the inclusion of $a_\mu$ in the fitting procedure could help reach the
$10^{-9}$ level, while hardly affecting the mechanism discussed in section \ref{sec:interpretation}. We observe nevertheless that the 
theoretical and experimental status of the $(g-2)_\mu$ tension still remains to be settled.

Concerning the WMAP constraint on the dark matter relic density, again we did not include this observable in our numerical analysis, since we 
consider it to be a very indirect constraint which can be spoiled by many cosmological effects. A posteriori, and under the assumption of a 
completely ``standard'' thermal history of the universe, we found values of the relic density much higher than allowed. 
However, it may be possible to sufficiently suppress the predicted relic density in the case it were included in the fitting procedure. In 
fact, since the lightest neutralino is always at the level of roughly 60 GeV, it might be possible, albeit through some fine tuning, to 
exploit the Higgs funnel region occurring at $2 \, m_{\chi^0_1} \simeq m_{h}$. For a detailed study of this possibility, see ref. \cite{Baer:2008jn}.

\section{Model-building discussion}\label{sec:SUSYbreaking}

The pattern of soft SUSY-breaking terms considered in this paper points to mechanisms of SUSY-breaking that do not, in general, 
respect the GUT group, and that, on the other hand, have a highly specific flavor structure. For example, in the general case of GUT-scale
MFV, the only flavor spurions generated should be the SM Yukawas. These mechanisms would be at work at or above the unification scale $M_G$.

As discussed in section \ref{sec:YU}, the trilinear splitting scenario of eqs. (\ref{eq:ST-m2})--(\ref{eq:ST-A}) analysed in this paper
is a special case of MFV soft terms. As illustrated in fig. \ref{fig:contours-inp}, for $m_{16} = 4$ TeV, our fitting procedure to 
low-energy data points to the following patterns of soft-breaking terms:
\beqn
&& \ov A_U \approx -2.5 \, m_{16}~, ~~~~ 0 \leq |\ov A_D| < |\ov A_U|~, \nn \\
\label{pattern}
&& m_{H_u}^2, |m_{H_d}^2| = {\rm O}(m_{16}^2)~,\\ 
&& m_{1/2} \ll m_{16}~.\nn
\eeqn
In this section we would like to discuss a concrete example of a SUSY-breaking scenario where this pattern is naturally realized.
Our example aims at illustrating to which extent, in a very predictive framework like that of Yukawa unification, the combined 
information from {\em existing} low-energy data can be translated into information on the mechanism of SUSY breaking at work.

The one pattern most clearly emerging from our analysis is a simultaneous splitting in $m_{H_u}^2 - m_{H_d}^2$ and in $\ov A_U - \ov A_D$.
This hints at SUSY-breaking VEVs coupled to operators that distinguish the `up' and `down' directions in Higgs soft-terms and 
$A$-terms alike. 
The arguably simplest scenario where this can be realized within SO(10) is with a spurion field that gets a VEV in the adjoint 
representation of SU(2)$_R$, namely $V \sim (1, 1, {\bf 3})$ (recall, in this respect, that SO(10) contains the Pati-Salam 
group SU(4) $\times$ SU(2)$_L$ $\times$ SU(2)$_R$ \cite{Pati-Salam}). One can further assume the presence of a spurion field $X$, that 
is instead a singlet under SO(10). A linear combination of the $X$ and $V$ fields will then introduce $F$-term SUSY breaking of the form
$c_1 F_X + c_2 F_V$, with $F_V$ proportional to the $T_3$ generator of SU(2)$_R$, distinguishing the `top' from the `bottom' direction.

$A$-term splitting will then be simply generated by the superpotential term $W_A \propto \ov{\mc{Q}} (c_1 X + c_2 V) H \mc{Q}$, where 
$\mc{Q}$ is the left-handed Pati-Salam chiral supermultiplet, transforming as $\mc{Q} \sim ({\bf 4}, {\bf 2}, 1)$, and containing the MSSM
chiral supermultiplets $Q$ and $L$ (see \cite{Raby-PDG} for an introduction to the formalism). The field $H$ contains instead the MSSM 
$H_u$ and $H_d$ superfields and transforms as $H \sim (1, {\bf 2}, {\bf 2})$. After the breakdown of SUSY, the coupling $W_A$ will then 
generate $A$-terms of the form $\ov A_{U, D} \propto c_1 \< F_X \> \pm c_2 \< F_V \> / 2$, the overall coefficient being fixed by the $W_A$ 
normalization.

The above field content easily gives rise to split $m_{H_{u,d}}^2$-terms as well. In fact, operators of the form 
$K_H \propto \chi_1^\dagger \chi_2 H^* H$, will naturally be present in the K{\"a}hler potential, where the fields $\chi_i$ can be chosen 
as $X$ or $V$, with gauge indexes adjusted so as to satisfy invariance under the Pati-Salam group. Among these operators, those inducing a 
single power of $F_V$ will distinguish $H_u$ from $H_d$, thereby splitting the corresponding soft terms.

It is worth observing that all the mass scales generated through the described mechanism are of the order $|\< F_{V,X} \>^2| / M_{\rm Planck}^2$,
which, in turn, will be of the order the gravitino mass, namely the TeV scale.

The above mechanism will also, in general, produce soft terms for squark and slepton bilinears. In particular, there is no obvious symmetry 
argument by which the K{\"a}hler potential operators mentioned above for Higgs bilinears would not induce sfermion bilinear splittings as well. 
In our paper, sfermion bilinears have been assumed all degenerate to the value $m_{16}$. The absence of splittings in our case has been justified 
on purely phenomenological grounds, namely, within the trilinear splitting scenario, data do not require bilinear splittings as well. 
The possibility that the above splitting mechanism could practically be ineffective for squark and slepton bilinears may 
be justified in frameworks where SUSY is broken through orbifold compactifications, and Higgs superfields live in the bulk whereas quark and 
lepton superfields are localized on the PS brane and the spurions $X$ and $V$ live on the SO(10) brane (see \cite{5DSO10} for an example). 
Finally, since $m_{1/2}$ is generated from the gauge kinetic function, it is completely unrelated to the above mechanism and can well be smaller 
in magnitude than $m_{16}$, as in eq. (\ref{pattern}).

The discussion in this section also highlights the importance of information on the lightest part of the SUSY spectrum, e.g. from the 
LHC. If the pattern were compatible with that described in section \ref{sec:spectrum}, this would allow, depending on, say, $M_{\tilde g}$ vs 
$M_{\tilde b_1}$ (see respectively figs. \ref{fig:contours-obs}, right and \ref{fig:contours-obs2}, left) to virtually select a point in the 
$A_b$ vs $\mu$ plane (see the panels of fig. \ref{fig:contours-inp}). 
The implied information on the main input parameters would permit correspondingly sharper model-building considerations than those 
presented above.

\section{Conclusion}\label{sec:conclusion}

We have considered general SUSY GUT frameworks with exact Yukawa unification and where the hypothesis of universal GUT-scale soft
terms is relaxed. We have first entertained the general possibility that soft terms be of minimally flavor violating form. In this case 
the hypothesis of exact YU and the hierarchical structure of the Yukawa couplings allow to parameterize squark bilinears
and trilinears in a general, simple and accurate way as in eqs. (\ref{eq:softMFVYU-m2})--(\ref{eq:softMFVYU-A}). Among the soft-term 
non-universalities allowed by this general parameterization, we have then focused on the scenario where up-type and down-type trilinear 
soft terms are split from each other.

We have explored the viability of this trilinear splitting scenario by contrasting the model predictions for EW observables, quark masses
and quark FCNC processes against data in a global fitting procedure. Agreement with data singles out one main scenario, featuring a sizable
splitting between the $A$-terms and a large $\mu$-term. In spite of a slight increase in the fine tuning required to achieve EWSB with
precisely the correct value of $M_Z$, this scenario allows a substantial improvement on other observables that, on a model-dependent basis,
do often require some amount of fine tuning as well.
First, and quite remarkably, phenomenological viability does not invoke a partial decoupling of the sparticle spectrum, as in the case of universal
soft terms, but instead it {\em requires} part of the spectrum, notably the lightest stop, the gluino and the lightest chargino and neutralinos,
to be very close to the current experimental limits. The lightest Higgs particle is also well above the LEP bound, it is actually quite robustly
predicted at around 125 GeV. Second, the above parameter space is selected by a non-trivial interplay between the requirement of negative,
sizable SUSY threshold corrections to $m_b$, and an instead negligible modification of the $B \to X_s \gamma$ decay rate, in presence of various
other EW and $B$-physics constraints. Hence the very same mechanism that makes the $m_b$ correction large enough, automatically
allows all FCNC constraints to be fulfilled.

We have also discussed a possible model of SUSY breaking where the pattern of soft terms, selected above on sheer phenomenological grounds,
is realized. This discussion highlights the crucial role of SUSY spectrum determinations at the LHC for either falsifying YU or else offering
important hints on the mechanism of SUSY breaking at work.

Our results provide a concrete example where, exploiting the full predictive power of YU and under what we consider very plausible 
assumptions for soft terms, enough remnant information on the high-energy symmetries does indeed survive at low energies for these symmetries 
to be reconstructible. They open up two natural directions of investigation. On the one side, a more complete exploration of the above discussed
general pattern of GUT-scale soft term non-universalities and its connection with a plausible mechanism of SUSY breaking. On the other side, 
a closer look at the collider signatures this pattern points to. Both directions are subject of future work.

\acknowledgments

D.G. owes special thanks to Gian Giudice, Gino Isidori and Giovanni Villadoro for insightful remarks. He also acknowledges Paride Paradisi 
for discussions and Stefan Recksiegel for instant help in making computational resources at TUM available. Finally,
the authors warmly thank Wolfgang Altmannshofer for making them available his FCNC library code.
The work of D.G. and D.M.S. has been supported in part by the Cluster of Excellence ``Origin and Structure of the Universe'' and by the 
German Bundesministerium f{\"u}r Bildung und Forschung under contract 05HT6WOA. D.G. also warmly acknowledges the support of the A. von 
Humboldt Stiftung. S.R. acknowledges partial funding from DOE grant DOE/ER/01545-882.

\bibliographystyle{JHEP}
\bibliography{GRS}

\end{document}